\documentclass[12pt]{article}
\usepackage{graphicx}
\usepackage{geometry}
\usepackage[applemac]{inputenc}

\usepackage{lmodern}
\usepackage{amsmath}
\usepackage{amsfonts}
\usepackage{amssymb}
\usepackage{amsthm}

\usepackage{natbib}

\usepackage[tight]{subfigure}
\ifx\pdfoutput\@undefined\usepackage[usenames,dvips]{color}
\else\usepackage[usenames,dvipsnames]{color}
\IfFileExists{pdfcolmk.sty}{\usepackage{pdfcolmk}}{} 
\fi
\usepackage[plainpages=false,pdfpagelabels,pagebackref=false,naturalnames=true,hyperindex=true,pdftitle={Information Measures of Complexity, Emergence, Self-organization, Homeostasis, and Autopoiesis},pdfauthor={N Fernandez, C. Maldonado, C. Gershenson}]{hyperref}
\hypersetup{colorlinks=true,
urlcolor=Cerulean,linkcolor=BrickRed,citecolor=RoyalBlue,
  pdfpagemode=None,
  pdfstartview=FitH}
\usepackage[all]{hypcap}

\usepackage{booktabs} 
\usepackage[table]{xcolor}
\usepackage{longtable}

\begin{document}

\title{Information Measures of Complexity, Emergence, Self-organization, Homeostasis, and Autopoiesis}
\author{ Nelson Fern\'{a}ndez$^{1,2}$, Carlos Maldonado$^{3,4}$ \&  Carlos Gershenson$^{4,5}$\\
$^{1}$ Laboratorio de Hidroinform\'{a}tica, Facultad de Ciencias B\'{a}sicas\\
Univesidad de Pamplona, Colombia\\
\url{http://unipamplona.academia.edu/NelsonFernandez}\\
$^{2}$ Centro de Micro-electr\'onica y Sistemas Distribuidos,\\
Universidad de los Andes, M\'erida, Venezuela\\
$^{3}$ Facultad de Ciencias \\
Universidad Nacional Aut\'onoma de M\'exico\\
$^{4}$ Departamento de Ciencias de la Computaci\'on\\
Instituto de Investigaciones en Matem\'aticas Aplicadas y en Sistemas \\
Universidad Nacional Aut\'onoma de M\'exico\\
$^{5}$ Centro de Ciencias de la Complejidad \\
Universidad Nacional Aut\'onoma de M\'exico\\
\href{mailto:cgg@unam.mx}{cgg@unam.mx} \
\url{http://turing.iimas.unam.mx/~cgg}
}
\maketitle

\begin{abstract}
This chapter reviews measures of emergence, self-organization, complexity, homeostasis, and autopoiesis based on information theory. These measures are derived from proposed axioms and tested in two case studies: random Boolean networks and an Arctic lake ecosystem. 

Emergence is defined as the information a system or process produces. Self-organization is defined as the opposite of emergence, while complexity is defined as the balance between emergence and self-organization. Homeostasis reflects the stability of a system. Autopoiesis is defined as the ratio between the complexity of a system and the complexity of its environment. The proposed measures can be applied at different scales, which can be studied with multi-scale profiles.
\end{abstract}

\section{Introduction}

In recent decades, the scientific study of complex systems~\citep{Bar-Yam1997,Mitchell:2009} has demanded a paradigm shift in our worldviews~\citep{GershensonEtAl-PnC,HeylighenEtAl2007}. Traditionally, science has been reductionistic. Still, complexity occurs when components are difficult to separate, due to relevant \emph{interactions}. These interactions are relevant because they generate novel information which determines the future of systems. This fact has several implications~\citep{Gershenson:2011e}. A key implication: reductionism---the most popular approach in science---is not appropriate for studying complex systems, as it attempts to simplify and separate in order to predict. Novel information generated by interactions limits prediction, as it is not included in initial or boundary conditions. It implies computational irreducibility~\citep{Wolfram:2002}, i.e.\ one has to reach a certain state before knowing it will be reached. In other words, \emph{a priori} assumptions are of limited use, since the precise future of complex systems is known only \emph{a posteriori}. This does not imply that the future is random, it just implies that the degree to which the future can be predicted is inherently limited.

It can be said that this novel information is \emph{emergent}, since it is not in the components, but produced by their interactions. Interactions can also be used by components to \emph{self-organize}, i.e.\ produce a global pattern from local dynamics. Interactions are also key for feedback control loops, which help systems regulate their internal states, an essential aspect of living systems.

We can see that reductionism is limited for describing such concepts as complexity, emergence, self-organization, and life. In the wake of the fall of reductionism as a dominant worldview~\citep{Morin2006}, a plethora of definitions, notions, and measures of these concepts has been proposed. Still, their diversity seems to have created more confusion than knowledge. In this chapter, we revise a proposal to ground measures of these concepts in information theory. This approach has several advantages: 
\begin{itemize}
\item Measures are precise and formal.
\item Measures are simple enough to be used and understood by people without a strong mathematical background.
\item Measures can help clarify the meaning of the concepts they describe.
\item Measures can be applied to any phenomenon, as anything can be described in terms of information~\citep{Gershenson:2007}. 
\end{itemize}

This chapter is organized as follows: In the next section, background concepts are presented, covering briefly complexity, emergence, self-organization, homeostasis, autopoiesis, information theory, random Boolean networks, and limnology. Section \ref{sec:measures} presents axioms and derives measures for emergence, self-organization, complexity, homeostasis and autopoiesis. To illustrate the measures, these are applied to two case studies in Section~\ref{sec:results}: random Boolean networks and an Arctic lake ecosystem. Discussion and conclusions close the chapter.

\section{Background}

\subsection{Complexity}

There are dozens of notions and measures of complexity, proposed in different areas with different purposes~\citep{Edmonds:1999,lloyd2001measures}. Etymologically, complexity comes from the Latin \emph{plexus}, which means interwoven. Thus, something complex is difficult to separate. This means that its components are interdependent, i.e.\ their future is partly determined by their \emph{interactions}~\citep{Gershenson:2011e}. Thus, studying the components in isolation---as reductionistic approaches attempt---is not sufficient to describe the dynamics of complex systems. 

Nevertheless, it would be useful to have global measures of complexity, just as temperature characterizes the properties of kinetic energy of molecules or photons. Each component can have a different kinetic energy, but the statistical average is represented in the temperature. For complex systems, particular interactions between components can be different, but we can say that complexity measures should represent the type of interactions between components, just as Lyapunov exponents characterize different dynamical regimes. 

A useful measure of complexity should enable us to answer questions such as: Is a desert more or less complex than a tundra? What is the complexity of different influenza outbreaks? Which organisms are more complex: predators or preys; parasites or hosts; individual or social? What is the complexity of different music genres? What is the required complexity of a company to face the complexity of a market\footnote{This question is related to the law of requisite variety~\citep{Ashby1956}.}?

Moreover, with the recent scandalous increase of data availability in most domains, we urgently need measures to make sense of it.

\subsection{Emergence}

Emergence has probably been one of the most misused concepts in recent decades. The reasons for this misuse are varied and include: polysemy (multiple meanings), buzzwording, confusion, hand waving, Platonism, and even mysticism. Still, the concept of emergence can be clearly defined and understood~\citep{Anderson1972}. The properties of a system are emergent if they are not present in their components. In other words, global properties which are produced by local interactions are emergent. For example, the temperature of a gas can be said to be emergent~\citep{Shalizi2001}, since the molecules do not possess such a property: it is a property of the collective. In a broad an informal way, emergence can be seen as differences in phenomena as they are observed at different scales~\citep{ProkopenkoEtAl2007}.

Some might perceive difficulties in describing phenomena at different scales~\citep{Gershenson:2011e}, but this is a consequence of attempting to find a single ``true" description of phenomena. Phenomena do not depend on the descriptions we have of them, and we can have several different descriptions of the same phenomenon. It is more informative to handle several descriptions at once, and actually it is necessary when studying emergence and complex systems. 

\subsection{Self-organization}

Self-organization has been used to describe swarms, flocks, traffic, and many other systems where the local interactions lead to a global pattern or behavior~\citep{CamazineEtAl2003,GershensonDCSOS}. 
Intuitively, self-organization implies that a system increases its own organization. This leads to the problems of defining organization, system, and self. Moreover, as Ashby showed~\citeyearpar{Ashby1947sos}, almost any dynamical system can be seen as self-organizing: if it has an attractor, and we decide to call that attractor ``organized", then the system dynamics will tend to it, thus increasing by itself its own organization. 
If we can describe almost any system as self-organizing, the question is not whether a system \emph{is} self-organizing or not, but rather, when is it useful to describe a system as self-organizing~\citep{GershensonHeylighen2003a}?

In any case, it is convenient to have a measure of self-organization which can capture the nature of local dynamics at a global scale. This is especially relevant for the nascent field of guided self-organization (GSO)~\citep{Prokopenko:2009,Ay2012Guided-self-org,GSO2013}.  GSO can be described as \emph{the steering of the self-organizing dynamics of a system towards a desired configuration}~\citep{Gershenson:2010}. This desired configuration will not always be the natural attractor of a controlled system. The mechanisms for guiding the dynamics and the design of such mechanisms will benefit from measures characterizing the dynamics of systems in a precise and concise way.

\subsection{Homeostasis}

Originally, the concept of homeostasis was developed to describe internal and physiological regulation of bodily functions, such as temperature or glucose levels. Probably the first person to recognize the internal maintenance of a near-constant environment as a condition for life was Bernard~\citeyearpar{Bernard1859}. Subsequently, Canon~\citeyearpar{Cannon:1932} coined the term homeostasis from the Greek h\'{o}moios (similar) and stasis (standing still). Cannon defined homeostasis as the ability of an organism to maintain steady states of operation during internal and external changes. Homeostasis does not imply an immobile or a stagnant state. Although some conditions may vary, the main properties of an organism are maintained.

Later, the British cybernetician William R. Ashby proposed, in an alternative form, that homeostasis implicates an adaptive reaction to maintain ``essential variables" within a range~\citep{Ashby1947,Ashby:1960}. In order to explain the generation of behavior and learning in machines and living systems, Ashby also contributed by linking the concepts of ultrastability and homeostatic adaptation~\citep{DiPaolo2000}. Ultrastability refers to the normal operation of the system within a ``viability zone" to deal with environmental changes. This viability zone is defined by the lower and upper bounds of the essential variables.  If the value of variables crosses the limits of its viability zone, the system has a chance of finding new parameters that make the challenged variables return to their viability zone.  

A dynamical system has a high homeostatic capacity if it is able to maintain its dynamics close to a certain state or states (attractors). As explained above, when perturbations or environmental changes occur, the system adapts to face the changes within the viability zone, that is, without the system ``breaking"~\citep{Ashby1947}. Homeostasis can be seen as a dynamic process of self-regulation and adaptation by which systems adapt their behavior over time~\citep{Williams2006}. The homeostasis concept can be applied to different fields beyond life sciences and is also closely related to self-organization and to robustness~\citep{Wagner2005,Jen2005}.

\subsection{Autopoiesis}

Autopoiesis comes from the Greek \emph{auto} (self) and \emph{poiesis} (creation, production) and was proposed as a concept to define the living. 
According to Maturana~\citeyearpar{Maturana2011}, the notion of autopoiesis was created to connote and describe the molecular processes taking place in the realization of living beings as autonomous entities. However, this meaning of the word autopoiesis, which was used to describe closed networks of molecular production, was chosen only until 1970~\citep{Maturana1980}. This notion arises from a series of questions, related to the internal dynamics of living systems, which Maturana began considering in the 1960s, such as: ``What should be the constitution of a system so that I see a living system as a result of its operation?", ``What kind of systems or entities are living systems?",  and another question that a student asked Maturana: ``What happened three billion eight hundred million years ago so that you can now say that living systems began then?"

In the context of autopoiesis, living beings occur as discrete autonomous dynamic molecular autopoietic entities. These entities are in a continuous realization of their self-production. Thus, autopoiesis describes the internal dynamics of a living system in the molecular domain. Maturana notices that living beings are dynamical systems in continuous change. Interactions between elements of an autopoietic system regulate the production and regeneration of the system’s components, having the potential to develop,  preserve, and produce their own organization \citep{varela1974autopoiesis}.

For example, a bacterium may produce another bacterium by cellular division, while a virus requires a host cell to produce another virus. The production of the new bacterium is made by the interactions between the elements of another bacterium. The production of a new virus depends on interactions between elements of an external system.
Thus, it can be said that a bacterium is more autopoietic than a virus.
In this sense, autopoiesis is much related to autonomy~\citep{RuizMoreno2004}. Autonomy is always limited in open systems, as their states depend on environmental interactions. However, differences in autonomy can be clearly identified, just like in the previous example.


The concept of autopoiesis has been extended to other areas beyond biology~\citep{Luisi2003,Seidl2004Luhmanns-theory,Froese2010}, although no formal measure had been proposed so far.

\subsection{Information Theory}

Information has had a most interesting history~\citep{gleick2011information}.
Information theory was created by Claude Shannon in 1948 in the context of telecommunications. He analyzed whether it was possible to reconstruct data transmitted across a noisy channel. In his model, \textbf{information} is represented as a string $X=x_{0}x_{1}...$ where each $x_{i}$ is a symbol from a finite set of symbols $ \mathcal{A}$ called the \textbf{alphabet}. Moreover, each symbol in the alphabet has a given probability $P(x)$ of occurring in the string. Common symbols will have a high $P(x)$ while infrequent symbols will have a low $P(x)$.

Shannon was interested in a function to measure how much information a process ``produces". Quoting \citet{Shannon1948}\footnote{We replaced Shannon's $H$ for $I$.}:
 
 \begin{quote}
   Suppose we have a set of possible events whose probabilities of occurrence are $p_{1}, p_{2}, ... ,p_{n}$. These probabilities are known but that is all we know about the event that might occur. Can we find a measure of how much ``choice'' is involved in the selection of the event or how uncertain we are of the outcome?
If there is such a measure, say $ (p_{1}, p_{2}, ... ,p_{n})$ it is reasonable to require of it the following properties:
\begin{enumerate}
  \item \emph{I} should be continuous in each $p_{i}$.
  \item If all the $p_{i}$ are equal, $p_{i} = 1/n$, then \emph{I} should be a monotonic increasing function of $n$. With equally $n$ likely events there is more choice, or uncertainty, when there are more possible events.
  \item If a choice be broken down into two successive choices, the original \emph{I} should be the weighted sum of the individual values of \emph{I}. 
\end{enumerate}

\end{quote}
With these few \emph{axioms}, Shannon demonstrates that the only function \emph{I} satisfying the three above is of the form:
\begin{equation} 
I = -K \sum_{i=i}^{n} p_{i} \log p_{i}, 
\label{eq:I}
\end{equation}
where $K$ is a positive constant. 
   
For example, if we have a string `0001000100010001...', we can estimate $P(0)=0.75$ and $P(1)=0.25$, then $I=-(0.75 \cdot \log{0.75}+0.25 \cdot \log{0.25})$. If we use $K=1$ and a base 2 logarithm, then $I\approx0.811$.

Shannon used $H$ to describe information (we are using $I$) because he was thinking in the Boltzmann's \emph{H} theorem\footnote{The Boltzmann \emph{H} theorem is given in the thermodinamic context. It states that the entropy of an ideal gas increases in an irreversible process. This might be also the reason why he required the second property.} when he developed the theory. Therefore, he called equation \ref{eq:I} the entropy of the set of probabilities  $p_{1}, p_{2}, ... ,p_{n}$. In modern words, \emph{I} is a function of a random variable $X$.

The unit of information is the bit (\emph{bi}nary digi\emph{t}). One bit represents the information gained when a binary random variable becomes known. However, since equation~\ref{eq:I} is a sum of probabilities, Shannon's information is a unitless measure.

More details about information theory in general can be found in~\citet{Ash1990Information}, while a primer on information theory related to complexity, self-organization, and emergence is found in~\citet{ProkopenkoEtAl2007}.



\subsection{Random Boolean Networks}

Random Boolean networks (RBNs) are abstract computational models, originally proposed to study genetic regulatory networks~\citep{Kauffman1969,Kauffman1993}. However, being general models, their study and use has expanded beyond biology~\citep{AldanaEtAl2003,Gershenson2004c,Gershenson:2010}.

A RBN is formed by $N$ nodes linked by $K$ connections\footnote{This $K$ is different from the constant used in equation~\ref{eq:I}.}. Each node has a Boolean state, i.e.\ zero or one. The future state of each node is determined by the current states of the nodes that link to it and a lookup table which specifies how the update will take place. The connectivity (which nodes affect which) and the lookup tables (how nodes affect their states) are usually generated randomly for a network, but remain fixed during its dynamics. 

RBNs have been found to have three different dynamical regimes, which have been studied extensively~\citep{Gershenson2004c}:
\begin{description}
\item[Ordered. ]  Most nodes are static, RBNs are robust to perturbations.
\item[Chaotic. ] Most nodes are changing, RBNs are fragile to perturbations.
\item[Critical. ] Some nodes are changing, RBNs have adaptive potential.
\end{description}
Different parameters and properties determine the regime, which can be used to guide a particular RBN towards a desired regime~\citep{Gershenson:2010}. 

It can be said that the critical regime balances the robustness of the chaotic regime and the changeability of the chaotic regime. It has been argued that computation and life require this balance to be able to compute and adapt~\citep{Langton1990,Kauffman1993}.
 
RBNs will be used in Section~\ref{sec:resultsRBNs} to illustrate the measures proposed in the next section.

\subsection{Limnology}

\begin{figure}[htbp]
\begin{center}
      \includegraphics[width = .95\textwidth]{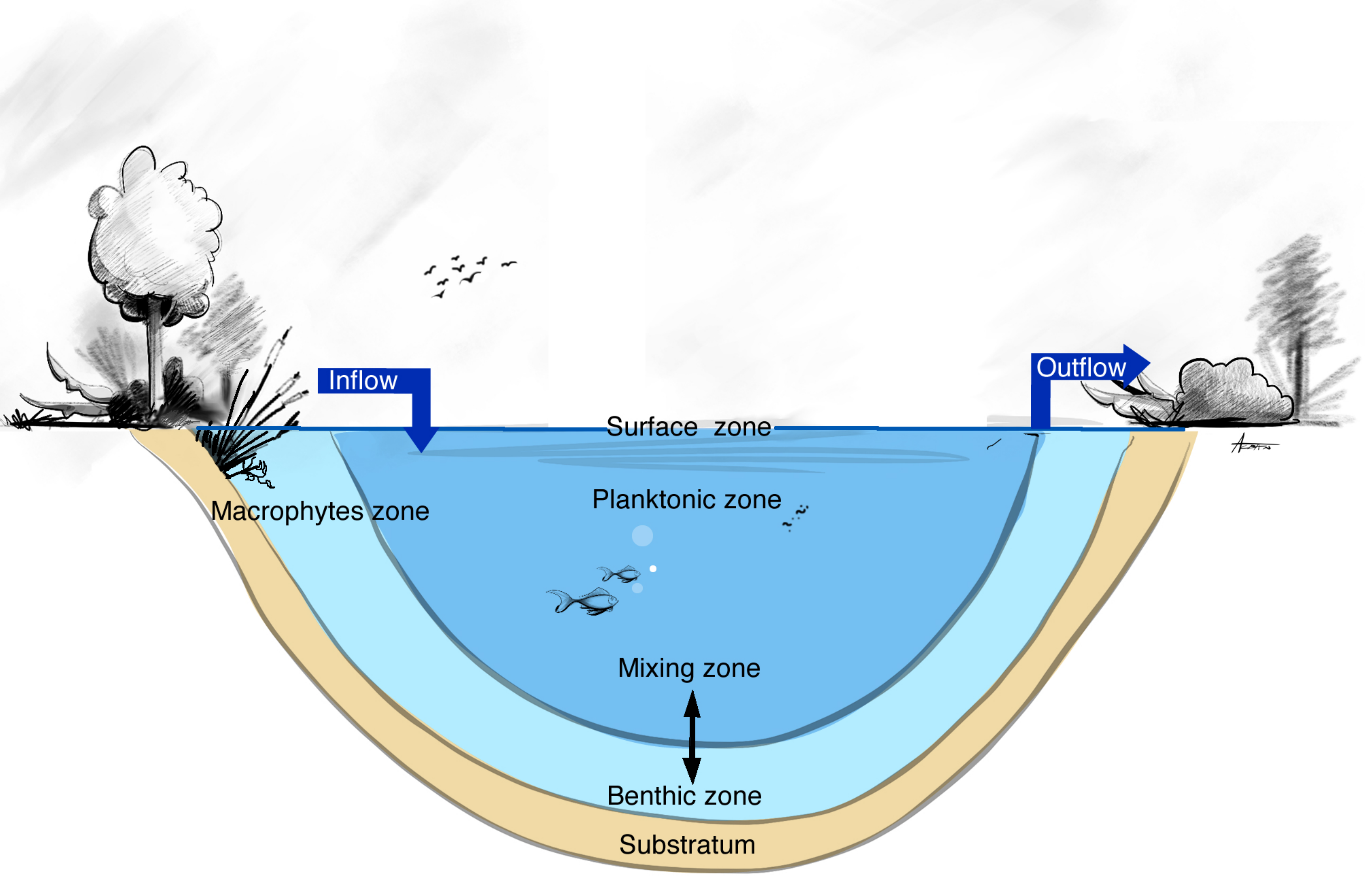}
\caption{Zones of lakes studied in limnology.}
\label{fig:lake}
\end{center}
\end{figure}

Lakes are studied by limnology. Lakes can be divided in different zones, as shown in Figure~\ref{fig:lake}: (i) The macrophyte zone, composed mainly of aquatic plants, which are rooted, floating or submerged. (ii) The planktonic  zone corresponds to the open surface waters; away from the shore in which organisms passively float and drift (phyto and zooplankton). Planktonic organisms are incapable of swimming against a current. However, some of them are motile. (iii) The benthic zone is the lowest level of a body of water related with the substratum, including the sediment surface and subsurface layers. (iv) The mixing zone is where the exchange of water from planktonic and benthic zones occurs.

At different zones, one or more components or subsystems can be an assessment for the ecosystem dynamics. For our case study to be presented in Section~\ref{sec:resultsLake}, we considered three components: physiochemical, limiting nutrients and photosynthetic biomass for the planktonic and benthic zones.

The physiochemical component refers to the chemical composition of water. It is affected by various conditions and processes such as geological nature, the water cycle, dispersion, dilution, solute and solids generation (e.g. photosynthesis), and sedimentation. In this component, we highlight two water variables that are important for the aquatic life: (i) the pH equilibrium that affects, among others, the interchange of elements between the organism and its environment and (ii) the temperature regulation that is supported in the specific heat of the water.

Related to the physiochemical component, limiting nutrients which are basic for photosynthesis  are associated with the biogeochemical cycles of nitrogen, carbon, and phosphorous.  These cycles permit the adsorption of gases into the water or the dilution of some limiting nutrients.

In addition, among limnetic biota, photoautotrophic biomass is the basis for the trophic web establishment. The term autotrophs is used for organisms that increase their mass through the accumulation of proteins which they manufacture, mainly from inorganic radicals~\citep{Stumm2004}. 
This type of organisms can be found at the planktonic and benthic zones.

The previous basic limnology concepts will be useful to follow the case study of an Arctic lake, presented in Section \ref{sec:resultsLake}.

\section{Measures}
\label{sec:measures}

We have recently proposed earlier versions of the measures presented in this chapter~\citep{Fernandez:2012,GershensonFernandez:2012}. The ones presented here are more refined and are based on axioms. The benefit of using axioms is that the discussion is not taken so much at the level of the measures, but at the level of the presuppositions or the properties we want measures to have.

A comparison of the proposed measures with others can be found in~\citet{GershensonFernandez:2012}. It is worth noting that all of the proposed measures are unitless.


\subsection{Emergence}
We mentioned that emergence refers to properties of a phenomenon which are present at one scale and are not at another scale. Scales can be temporal or spatial. If we describe phenomena in terms of information, in order to have ``new" information, ``old" information has to be \emph{transformed}. This transformation can be dynamic, static, active, or stigmergic~\citep{Gershenson:2007}. For example, new information is produced when a dynamical system changes its behavior, but also when a description of a system changes. Concerning the first case, approaches measuring the difference between past and future states have been proposed, e.g.~\citep{ShaliziCrutchfield2001}. We can call this \emph{dynamic emergence}. Concerning the second case, approaches measuring differences between scales have been used, e.g.~\citep{Shalizi2001,Holzer:2011}. We can call this \emph{scale emergence}. Even when there are differences between dynamic and scale emergencies, both can be seen as new information being produced. In the first case, dynamics produce new information. In the second case, the change of description produces new information. Thus, \emph{information emergence} $E$ includes both dynamic emergence and scale emergence.
If we recall, Shannon proposed a quantity which measures how much information a process ``produces". Therefore, we can say that emergence is the same as Shannon's information $I$. 
From now on, we will consider the emergence of a process \emph{E} as the information \emph{I} and we will use the base two logarithm. 

\begin{equation}
E=I.
\label{eq:E}
\end{equation}

We now revise that the intuitive idea of emergence fulfills the three basic notions (axioms) that Shannon used to derive \emph{I} (Shannon's $H$). For the continuity axiom, it is expected of a measure not to give big jumps when small changes are made. The second axiom will be harder to show. It states that if we consider an auxiliary function $i$ which is the \emph{I} function when there are $n$ events with the same probability $1/n$ then the function $i$ is monotonic increasing. If we have the same configuration for emergence, then we could think the process to be with equally likelihood in any of $n$ available states. If something happens and now the process can be in $ n + k $ equally likely states we can say that the process has had emergence, since now we need more information to know in which state the process is. 
For the third axiom, we need to find a way to figure out how is that we can 'split' the process. Lets recall that the third property required by Shannon is that if a choice can be broken into two different choices, the original \emph{I} should be the average of the other two \emph{I}. In a process, we can think the choices as a fraction of the process that we are currently observing. For this purpose, we can make a partition\footnote{We are using the set theory partition, we could have any finite number of partitions where the intersection of all of them is the null set and whose union is the original set.} of the domain, in our case, we get two subsets whose intersection is the null set and whose union is the full original set. After this, we compute the \emph{I} function for each. Since we observe two different parts of a process and in each observation we get the average\footnote{When there are more than two subsets in the partition, we can make a weighted average. A sort of expectation where the distribution probability is given by the nature of the process.} new information required to describe the (partial) process, then it makes sense to take the average of both when observing the full process. 



$E$, as well as $I$, is a probabilistic measure.
$E=1$ means that when any random binary variable becomes known, one bit of information emerges. If $E=0$, then no new information will emerge, even as random binary variables become ``known" (they are known beforehand).
Again, we emphasize that emergence can take place at the level of a phenomenon observed or at the level of the description of the phenomenon observed. Either can produce novel information. 

 \subsubsection{Multiple Scales}
 
When Shannon defined equation~\ref{eq:I}, he included $K$ which is a positive constant. This is important because we will change the value of $K$ to normalize a measure onto the $[0,1]$ interval. The value of $K$ will depend on the length of the finite alphabet $\mathcal{A}$ we use. In the particular Boolean case when we have the alphabet $\mathcal{A} = \{0,1\}$ with length $|\mathcal{A}|$ = 2. Then the value $K = 1$ will normalize the measure to the interval $[0,1]$. Because of the relevance of the binary notation in computer science and other applications, we will often use the Boolean alphabet. Nevertheless, we can compute the entropy for alphabets with different lengths. 
We only have to consider the equation

\begin{equation}
K = \frac{1}{\log_{2}b},
\end{equation}
where $b$ is the length of the alphabet we use. In this way we will normalize $E$ and measures derived from it, having a maximum of 1 and a minimum of 0. 

For example, consider the string in base 4 `0133013301330133...'. We can estimate $P(0)=P(1)=0.25$, $P(2)=0$, and $P(3)=0.5$. Following equation~\ref{eq:I}, we have 
$I=-K(0.25 \cdot \log{0.25} + 0.25 \cdot \log{0.25} + 0 + 0.5 \cdot \log{0.5})$. Since $b=4$, $K = \frac{1}{\log_{2}4}=0.5$. Thus, we obtain a normalized $I=0.75$.

\subsection{Self-organization}

    
Self-organization has been correlated with an increase in order, i.e.\ a reduction of entropy~\citep{GershensonHeylighen2003a}. If emergence implies an increase of information, which is analogous to entropy and disorder, self-organization should be anti-correlated with emergence.
   
 
  A measure of self-organization $S$ should be a function $S:\Sigma \rightarrow \mathbb{R}$ (where $\Sigma =  \mathcal{A}^\mathbb{N}$) with the following properties:
  \begin{enumerate}
  
  \item The range of $S$ is the real interval $ [ 0,1 ] $ 
  \item $S(X) = 1$ if and only if $X$ is deterministic, i.e.\ we know beforehand the value of the process. 
  \item $S(X) = 0$ if and only if $X$ has a uniform distribution, i.e.\ any state of the process is equally likely. 
  \item $S(X)$ has a negative correlation with emergence $E$.
  
  \end{enumerate}
   
    We propose as the measure 
\begin{equation}
S = 1 - I = 1 - E
\label{eq:S}
\end{equation}

 It is straightforward to check that this function fulfills the axioms stated. Nevertheless it is not unique. However, it is the only affine (linear) function which fulfills the axioms. For simplicity, we propose the use of~\ref{eq:S} as a measure of self-organization.

$S=1$ means that there is maximum order, i.e.\ no new information is produced ($I=E=0$). On the other extreme, $S=0$ when there is no order at all, i.e.\ when any random variable becomes known, information is produced/emerges ($I=E=1$). When $S=1$, maximum order, dynamics do not produce novel information, so the future is completely known from the past. On the other hand, when $S=0$, minimum order, no past information tells us anything about future information. 

Note that equation~\ref{eq:S} makes no distinction on whether the order is produced by the system (self) or by its environment. Thus, $S$ would have a high value in systems with a high organization, independently on whether this is a product of local interactions or imposed externally. This distinction can be easily made describing the detailed behavior of systems, but is difficult and unnecessary to capture with a general probabilistic measure such as $S$. As an analogy, one can measure the temperature of a substance, but temperature does not differentiate (and does not need to differentiate) between substances which are heated or cooled from the outside and substances whose temperature is dependent mainly on internal chemical reactions.
     
\subsection{Complexity}

Following \citet{LopezRuiz:1995}, we can define complexity $C$ as the balance between change (chaos) and stability (order). We have just defined such measures: emergence and self-organization. The complexity function $C:\Sigma \rightarrow \mathbb{R}$ should have the following properties:

\begin{enumerate}
\item The range is the real interval $[0,1]$.
\item $C = 1$ if and only if $S = E$.
\item $C = 0$ if and only if $S = 0$ or $E = 0$.  
\end{enumerate} 

It is natural to consider the product of $S$ and $I$ to satisfy the last two requirements. We propose:
\begin{equation} 
C = 4 \cdot E \cdot S. 
\label{eq:C}
\end{equation}
Where the constant 4 is added to normalize the measure to $[0,1]$, fulfilling the first axiom. $C$ can also be represented in terms of $I$ as:
\begin{equation} 
C = 4 \cdot I \cdot (1-I). 
\label{eq:C2}
\end{equation}

\begin{center}
\begin{figure}
   \begin{center}
      \includegraphics[width = .5\textwidth]{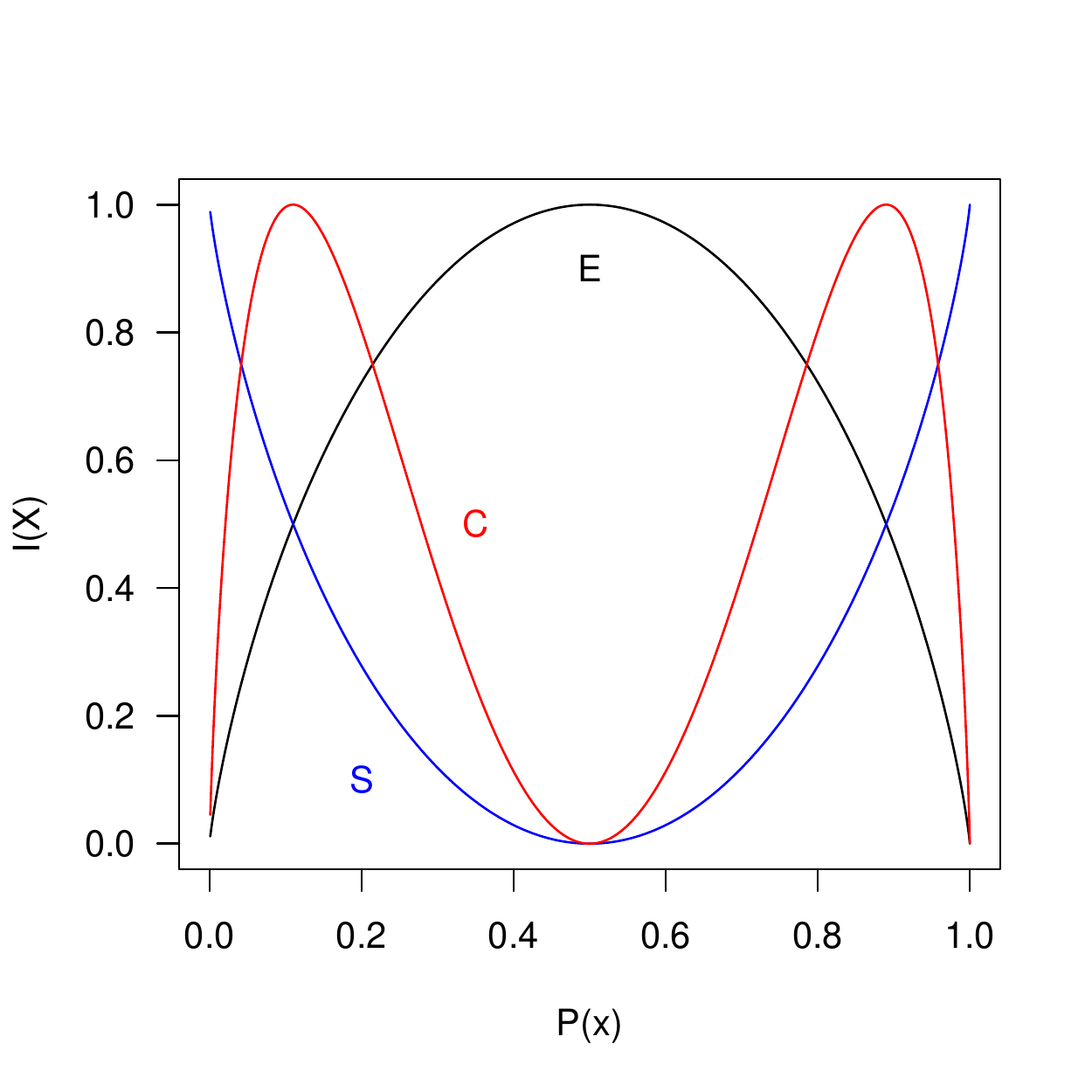}
      \label{fig:ESC}
      \caption{Emergence $E$, self-organization $S$, and complexity $C$.}
   \end{center}
\end{figure}
\end{center}

Figure~\ref{fig:ESC} plots the measures proposed so far for different values of $P(x)$. It can be seen that $E$ is maximal when $P(x)=0.5$ and minimal when $P(x)=0$ or $P(x)=1$. The opposite holds for $S$: it is minimal when $P(x)=0.5$ and maximal when $P(x)=0$ or $P(x)=1$. $C$ is minimal when $S$ or $E$ are minimal, i.e.\ $P(x)=0$, $P(x)=0.5$, or $P(x)=1$. $C$ is maximal when $E=S=0.5$, which occurs when $P(x) \approx 0.11$ or $P(x) \approx 0.89$.

Shannon information can be seen as a balance of zeros and ones (maximal when $P(0)=P(1)=0.5$), while $C$ can be seen as a balance of $E$ and $S$ (maximal when $E=S=0.5$).

\subsection{Homeostasis}

The previous three measures ($E$, $S$, and $C$) study how single variables change in time. To calculate the measures for a system, one can plot the histogram or simply average the measures for all variables in a system. For homeostasis $H$, we are interested on how all variables of a system change or not in time. Table~\ref{tab:variables} shows this difference: $E$, $S$, and $C$ focusses on time series of \emph{variables} (columns), while $H$ focusses on \emph{states} (rows).

\begin{table}[htdp]
\caption{Difference between observing single variables in time (columns) and several variables at one time (rows).}
\begin{center}
\begin{tabular}{|l|l|l|l|}
\hline
	&	$X$	&		$Y$	&		$Z$	\\
\hline
$t=m-2$			&	$x_{m-2}$&	$y_{m-2}$&	$z_{m-2}$\\				
$t=m-1$			&	$x_{m-1}$&	$y_{m-1}$&	$z_{m-1}$\\				
$t=m$			&	$x_{m}$&		$y_{m}$&		$z_{m}$\\				
\hline
\end{tabular}
\end{center}
\label{tab:variables}
\end{table}%

Let $X=x_1x_2x_3...x_n$ represent the \emph{state} of a system of $n$ variables (i.e.\ a row in Table~\ref{tab:variables}). If the system has a high homeostasis, we would expect that its states do not change too much in time. The homeostasis function $H:\Sigma \times \Sigma \rightarrow \mathbb{R}$ should have the following properties:

\begin{enumerate}
\item The range is the real interval $[0,1]$.
\item $H = 1$ if and only if for states $X$ and $X'$, $X = X'$, i.e.\ there is no change in time.
\item $H = 0$ if and only if $\forall i,  x_i \neq x'_i$, i.e.\ all variables in the system changed.  
\end{enumerate} 

A useful function for comparing strings of equal length is the Hamming distance. The Hamming distance $d$ measures the percentage of different symbols in two strings $X$ and $X'$. For binary strings, it can be calculated with the XOR function ($\oplus$). Its normalization bounds the Hamming distance to the interval $[0,1]$:
\begin{equation}
d(X,X')= \frac{\sum\limits_{i \in \{ 0, \dots ,|X|\} }{x_i \oplus x'_i}}{|X|}.
\label{eq:d}
\end{equation}
$d$ measures the fraction of different symbols between $X$ and $X'$. For the Boolean case, $d=0 \iff X=X'$ and $d=1 \iff X=\neg X'$, while $X$ and $X'$ are uncorrelated $\iff d\approx0.5$.

We can use the inverse of $d$ to define $h$:
\begin{equation}
h(X^t,X^{t+1})=1-d(X^t,X^{t+1}),
\label{eq:H1}
\end{equation}
which clearly fulfills the desired properties of homeostasis between two states.

To measure the homeostasis of a system in time, we can generalize:

\begin{equation}
H=\frac{1}{m-1}\sum_{t=0}^{m-1}h(X^t,X^{t+1}),
\label{eq:H}
\end{equation}

where $m$ is the total number of time steps being evaluated. $H$ will be simply the average of different $h$ from $t=0$ to $t=m-1$. As well as the previous measures based on $I$, $H$ is a unitless measure.

When $H$ is measured at higher scales, it can capture periodic dynamics. For example, let us have a system with $n=2$ variables and a cycle of period 2: $11 \rightarrow 00 \rightarrow 11$. $H$ for base 2 will be minimal, since every time step all variables change, i.e.\ ones turn into zeros or zeros turn into ones. However, if we measure $H$ in base 4, then we will be actually comparing pairs of states, since to make one time step in base 4 we take two binary time steps. Thus, in base 4 the attractor becomes $22 \rightarrow 22$, and $H=1$. The same applies for higher bases. An example of the usefulness of measuring $H$ at multiple scales in elementary cellular automata is explained in~\citet{GershensonFernandez:2012}.

\subsection{Autopoiesis}

Let $\bar{X}$ represent the trajectories of the variables of a system and $\bar{Y}$ represent the trajectories of the variables of the environment of the system. A measure of autopoiesis $A:\Sigma \times \Sigma \rightarrow \mathbb{R}$ should have the following properties:

\begin{enumerate} 
\item $A \geq 0$.
\item $A$ should reflect the independence of $\bar{X}$ over $\bar{Y}$. This implies:
\begin{enumerate}
\item $A>A' \iff \bar{X}$ produces more of its own information than $\bar{X'}$ for a given $\bar{Y}$.
\item $A>A' \iff \bar{X}$ produces more of its own information in $\bar{Y}$ than in $\bar{Y'}$.
\item $A=A' \iff \bar{X}$ produces as much of its own information than $\bar{X'}$ for a given $\bar{Y}$.
\item $A=A' \iff \bar{X}$ produces as much of its own information in $\bar{Y}$ than in $\bar{Y'}$.
\item $A=0$ if all of the information in $\bar{X}$ is produced by $\bar{Y}$.
\end{enumerate}
\end{enumerate} 

Following the classification of types of information transformation proposed in~\citet{Gershenson:2007}, dynamic and static transformations are internal (a system producing its own information), while active and stigmergic transformations are external (information produced by another system).

It is problematic to define in a general and direct way how some information depends on other information, as causality can be confounded with co-occurrence. For this reason, measures such as mutual information are not suitable for measuring $A$. 

As it has been proposed, adaptive systems require a high $C$ in order to be able to cope with changes of its environment while at the same time maintaining their integrity~\citep{Langton1990,Kauffman1993}. If $\bar{X}$ had a high $E$, then it would not be able to produce the same patterns for different $\bar{Y}$. With a high $S$, $\bar{X}$ would not be able to adapt to changes in $\bar{Y}$. Therefore, we propose:

\begin{equation}
A=\frac{C(\bar{X})}{C(\bar{Y})}.
\label{eq:A}
\end{equation}

If $C(\bar{X})=0$, then either $\bar{X}$ is static ($E(\bar{X})=0$) or pseudorandom ($S(\bar{X})=0$). This implies that any pattern (complexity) which could be observed in $\bar{X}$ (if any) should come from $\bar{Y}$.  This case gives a minimal $A$. On the other hand, if $C(\bar{Y})=0$, it implies that any pattern (if any) in $\bar{X}$ should come from itself. This case gives a maximal $A=\infty$. A particular case occurs if $C(\bar{X})=0$ and $C(\bar{Y})=0$. $A$ becomes undefined. But how can we say something about autopoiesis if we are comparing two  systems which are either without variations ($S=1$) or pseudorandom ($E=1$)? This case \emph{should} be undefined.
The rest of the properties are evidently fulfilled by equation~\ref{eq:A}. This is certainly not the unique function to fulfill the desired axioms. The exploration of alternatives requires further study.

Since $A$ represents a ratio of probabilities, it is a unitless measure. $A\in[0,\infty)$, although it could be mapped to $[0,1)$ using a function such as $f(A) = \frac{A}{1+A}$. We do not normalize $A$ because it is useful to distinguish $A>1$ and $A<1$ (see Section~\ref{sec:ReqVar}).

\subsection{Multi-scale profiles}

Bar-Yam~\citeyearpar{BarYam2004} proposed the ``complexity profile", which plots the complexity of systems depending on the scale at which they are observed. This allows to compare how a measure changes with scale. For example, the $\sigma$ profile compares the ``satisfaction" of systems at different scales to study organization, evolution and cooperation~\citep{Gershenson:2010a}.

In a similar way, multi-scale profiles can be used for each of the measures proposed, giving further insights about the dynamics of a system than measuring them at a single scale. This is clearly seen, for example, with different types of elementary cellular automata~\citep{GershensonFernandez:2012}.

\section{Results}
\label{sec:results}

In this section we apply the measures proposed in the previous section to two case studies: random Boolean networks and an aquatic ecosystem. A further case, elementary cellular automata, can be found in~\citet{GershensonFernandez:2012}.

\subsection{Random Boolean Networks}
\label{sec:resultsRBNs}

Results show averages of 1000 RBNs, where 1000 steps were run from a random initial state and $E$, $S$, $C$ and $H$ were calculated from data generated in 1000 additional steps.

R~\citep{R} was used with packages \emph{BoolNet}~\citep{Mussel:2010} and \emph{entropy}~\citep{Hausser:2012}.

\begin{figure}[htbp]
\begin{center}
  \includegraphics[width=0.95\textwidth]{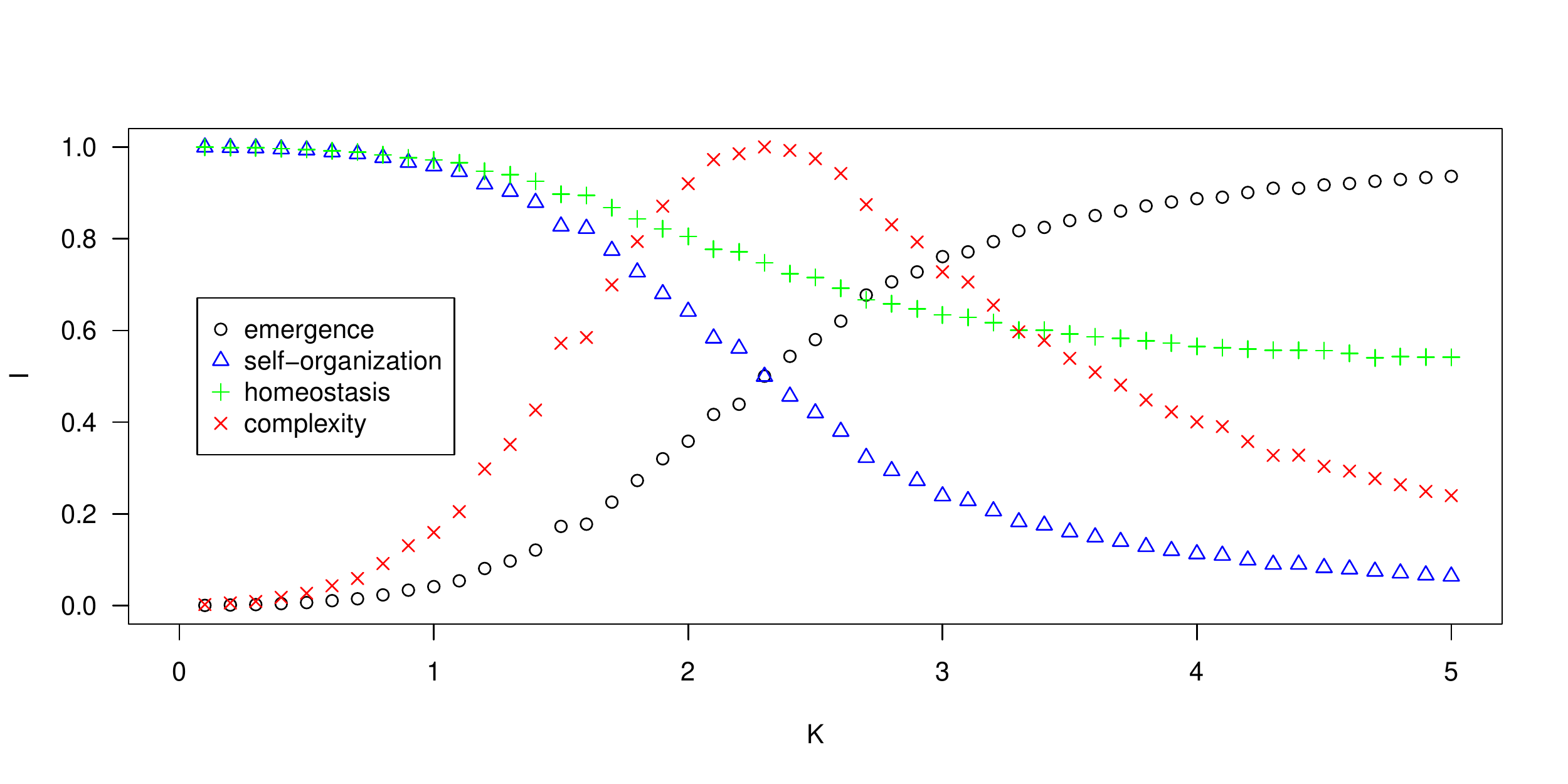}\\
\caption{Averages for 1000 RBNs, $N=100$ nodes and varying average connectivity $K$ \citep{GershensonFernandez:2012}.}
\label{InfoRBN-pl}
\end{center}
\end{figure}

Figure~\ref{InfoRBN-pl} shows results for RBNs with 100 nodes, as the connectivity $K$ varies. For low $K$, there is high $S$ and $H$, and a low $E$ and $C$. This reflects the ordered regime of RBNs, where there is high robustness and few changes. Thus, it can be said that there is few or no information emerging and there is a high degree of self-organization and homeostasis.
For high $K$, there is high $E$, low $S$ and $C$, and uncorrelated $H\approx0.5$. This reflects the chaotic regime of RBNs, where there is high fragility and many changes. Almost every bit (a new state for most nodes) carries novel emergent information, and this constant change implies low organization and complexity. For medium connectivities ($2\leq K \leq 3$), there is a balance between $E$ and $S$, leading to a high $C$. This corresponds to the critical regime of RBNs, which has been associated with complexity and the possibility of life~\citep{Kauffman2000}.

As for autopoiesis, to model a system and its environment, we coupled two RBNs: One ``internal" RBN with $N_i$ nodes and $K_i$ average connections and one ``external" with $N_e$ nodes and $K_e$ average connections. A ``coupled" RBN is considered with $N_c=N_i+N_e$ nodes and $K_i$ connections. At every time step, the external RBN evolves independently. However, its state is copied to the $N_e$ nodes representing it in the coupled RBN, which now evolves depending partly on the external RBN. Thus, the $N_i$ nodes in the coupled RBN representing the internal RBN may be affected by the dynamics of the external RBN, but not vice versa. The $C$ of each node is calculated and averaged separately for each network, obtaining an internal complexity $C_i$ and an external complexity $C_e$.

\begin{figure}[htbp]
\begin{center}
  \includegraphics[width=0.75\textwidth]{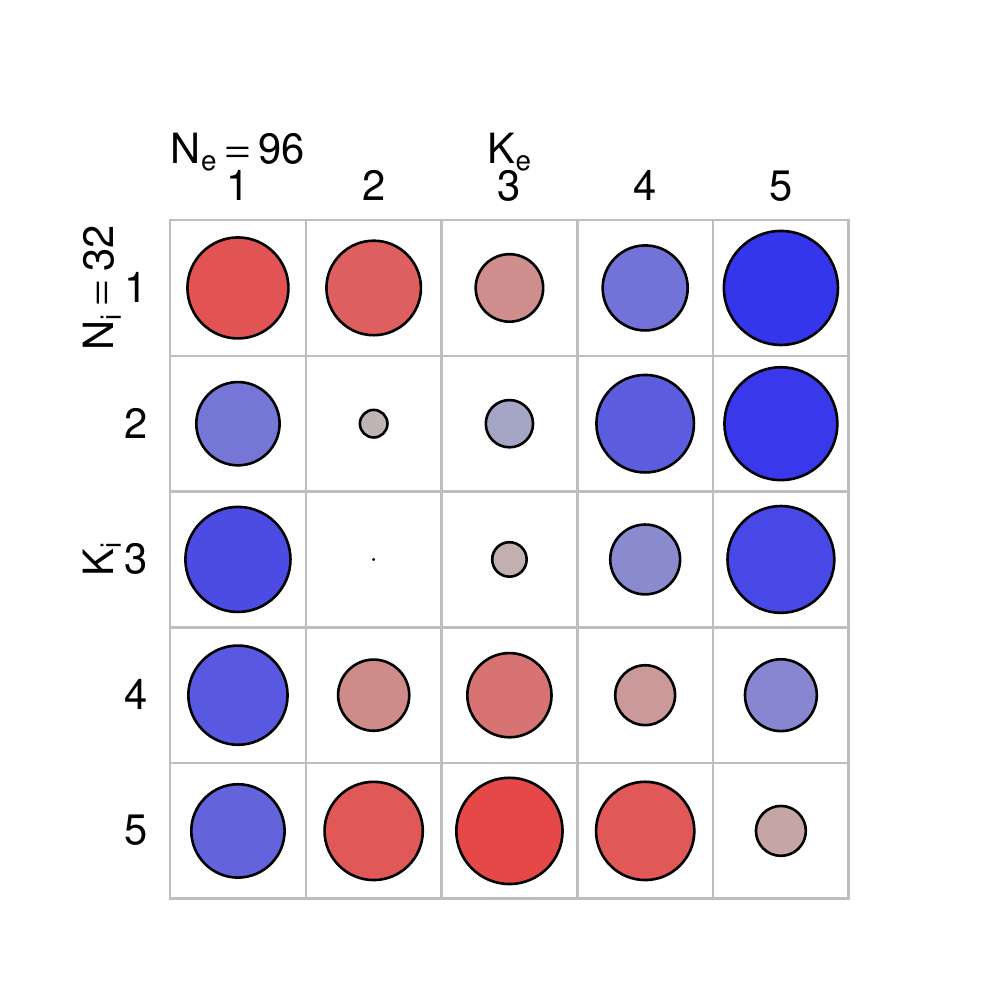}\\
\caption{$A$ averages for 50 sets $N_e=96, N_i=32$. Values $A<1$ are red while $A>1$ are blue. Size of circles indicate how far $A$ is from $A=1$. Numerical values shown in Table~\ref{tab:RBN-A}.}
\label{fig:RBN-A}
\end{center}
\end{figure}

Figure~\ref{fig:RBN-A} and Table~\ref{tab:RBN-A} show results for $N_e=96$ and $N_i=32$ for different combinations of $K_e$ and $K_i$.

\begin{table}[htdp]
\caption{$A$ averages for 50 sets $N_e=96, N_i=32$. Same results as those shown in Figure~\ref{fig:RBN-A}.}
\begin{center}
\begin{tabular}{|c|c|c|c|c|c|}
\hline
$K_i$ \textbackslash $K_e$ &	1&	2&	3&	4&	5 \\
\hline
1&	0.4464025& 0.5151070& 0.7526248& 1.6460345& 3.4081967\\
2&	1.6043330& 0.9586809& 1.1379227& 2.0669794& 3.2473729\\
3&	 2.4965328& 0.9999926& 0.9355231& 1.3604272& 2.6283798\\
4&	2.1476247& 0.7249803& 0.6151742& 0.8055051& 1.3890630\\
5&	1.8969094& 0.4760027& 0.3871875& 0.4755580& 0.8648389\\
\hline
\end{tabular}
\end{center}
\label{tab:RBN-A}
\end{table}%

As it was shown in Figure~\ref{InfoRBN-pl}, $C$ changes with $K$, so it is expected to have $A\approx 1$ when $K_i\approx K_e$. When $C_e$ is high ($K_e=2$ or $K_e=3$), then the environment dominates the patterns of the system, yielding $A<1$. When $C_e$ is low ($K_e<2$ or $K_e>3$), the patterns produced by the system are not affected that much by its environment, thus $A>1$, as long as $K_i<K_e$ (otherwise the system is more chaotic that its environment, and so complex patterns have to come from outside).

$A$ does not try to measure how much information emerges internally or externally, but how much the patterns are internally or externally produced. A high $E$ means that there is no pattern, as there is constant change. A high $S$ implies a static pattern. A high $C$ reflects complex patterns. We are interested in $A$ measuring the ratio of the complexity of patterns being produced by a system \emph{compared} to the complexity of patterns produced by its environment.

\subsection{An Ecological System: An Arctic Lake}
\label{sec:resultsLake}

The data from an Artic lake model used in this section was obtained using The Aquatic Ecosystem Simulator~\citep{Randerson2008}.

\begin{figure}
     \centering
     \subfigure[]{
          \label{fig:temp}
          \includegraphics[width=.9\textwidth]{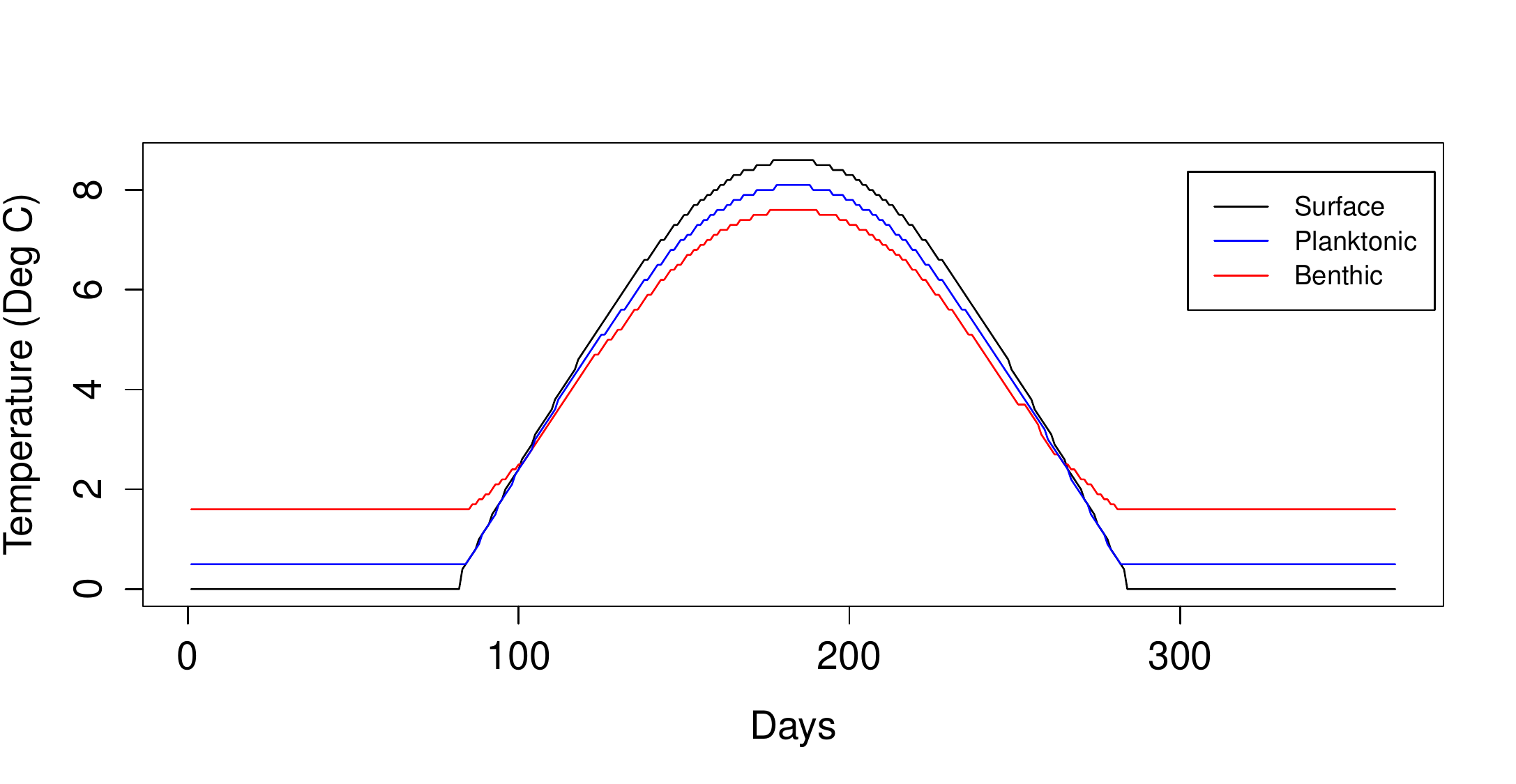}}\\
          
     \subfigure[]{
          \label{fig:hydro}
          \includegraphics[width=.9\textwidth]{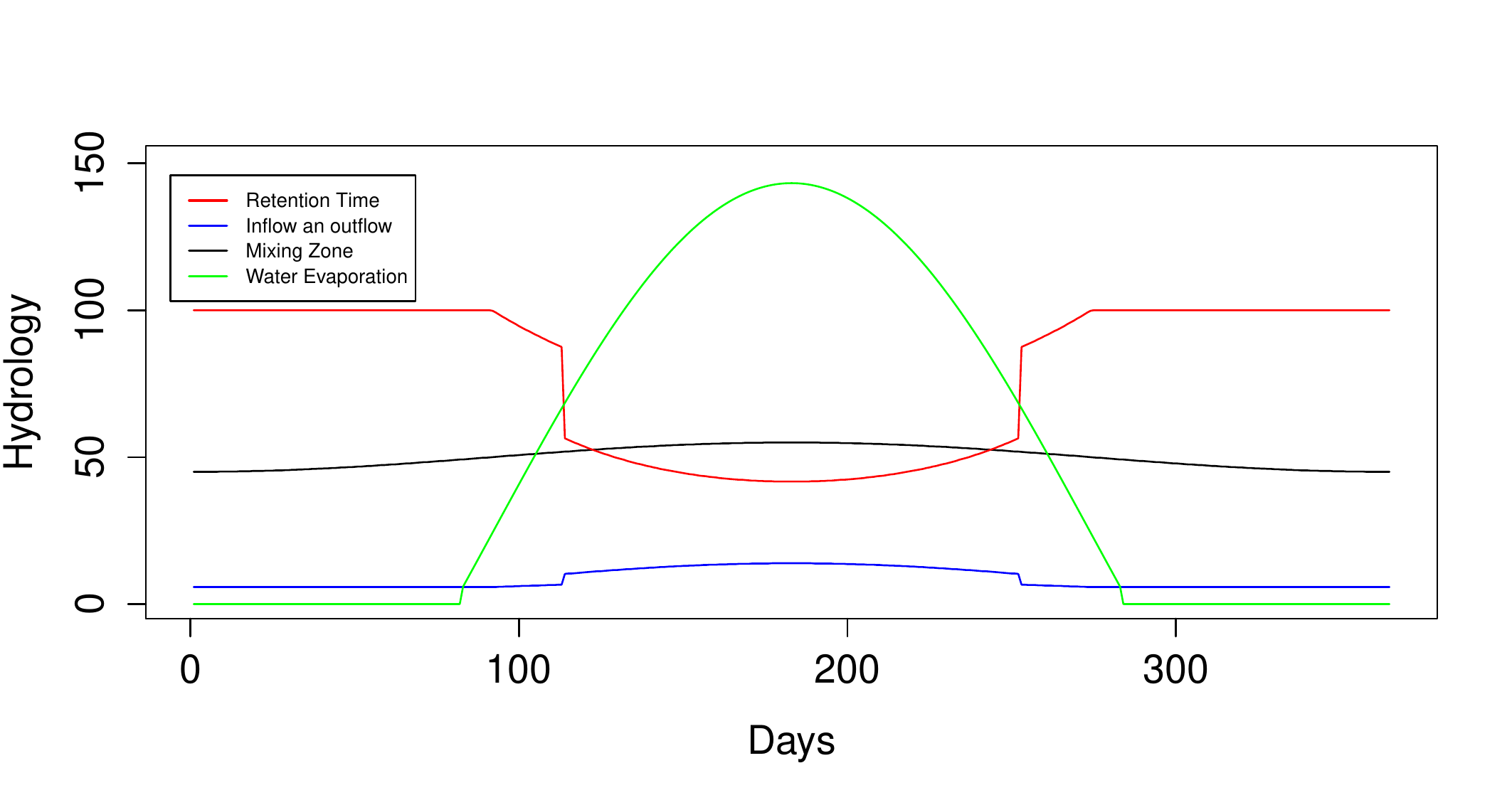}}
     \caption{(A) Climatic and (B) hydraulic regimes of Arctic lakes.}
     \label{fig:regimes}
\end{figure}

In general, Arctic lake systems are classified as oligotrophic due to their low primary production, represented in chlorophyll values of 0.8-2.1 mg/m3. The lake’s water column, or limnetic zone, is well-mixed; this means that there are no stratifications (layers with different temperatures). During winter (October to March), the surface of the lake is ice covered. During summer (April to September), ice melts and the water flow and evaporation increase, as shown in Figure~\ref{fig:temp}. Consequently, the two climatic periods (winter and summer) in the Arctic region cause a typical hydrologic behavior in lakes as the one shown in Figure~\ref{fig:hydro}. This hydrologic behavior influences the physiochemical subsystem of the lake. 

\begin{figure}[htbp]
\begin{center}
  \includegraphics[width=0.98\textwidth]{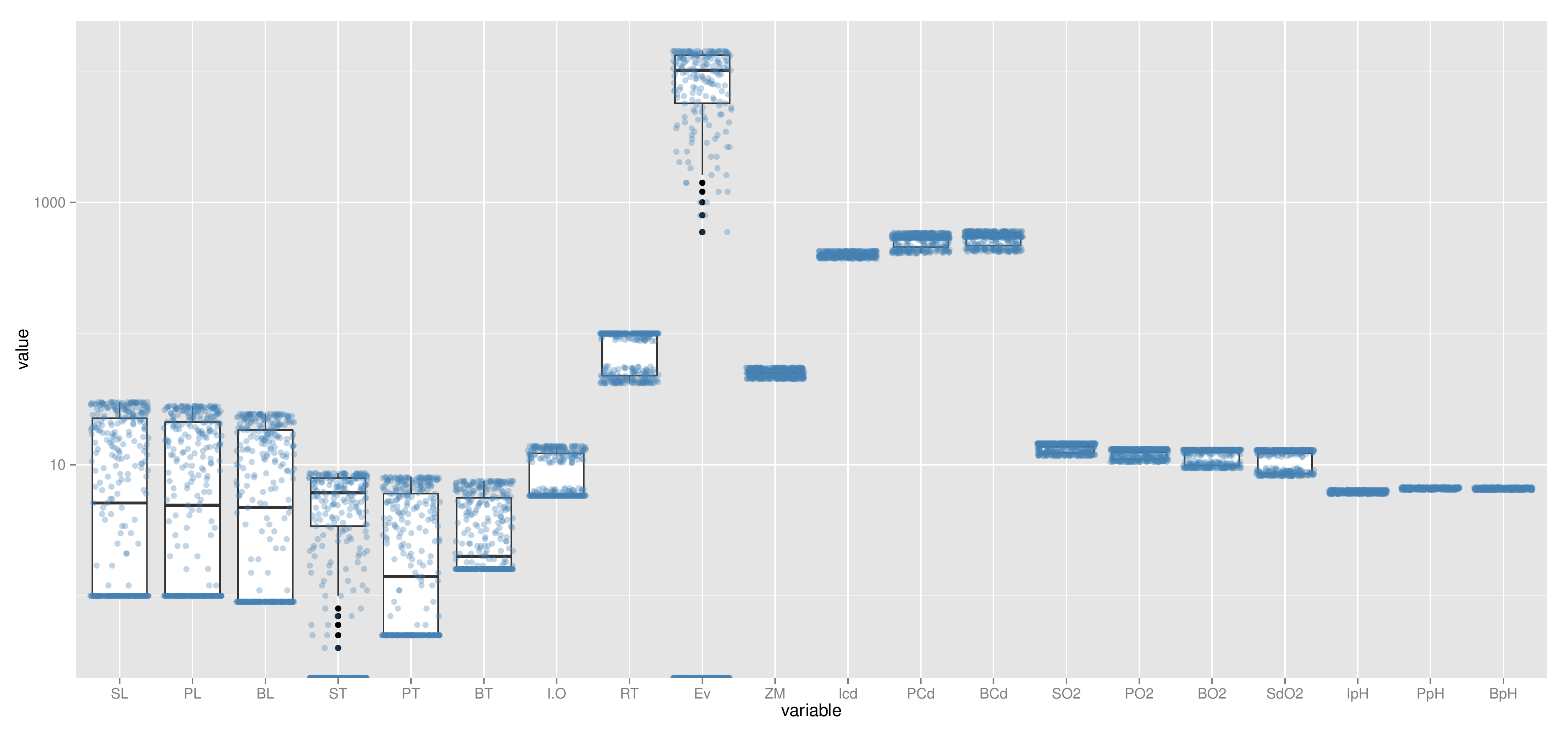}\\
\caption{Boxplots of variables from the physiochemical subsystem. Abbreviations expanded in Table~\ref{tab:vars}.}
\label{fig:vars}
\end{center}
\end{figure}

Table~\ref{tab:vars} and Figure~\ref{fig:vars} show the variables and daily data we obtained from the Arctic lake simulation. The model used is deterministic, so there is no variation in different simulation runs. Figure~\ref{fig:vars} depicts a high dispersion for the following variables: temperature ($T$) and light ($L$) at the three zones of the Arctic lake (surface=$S$, planktonic=$P$ and benthic=$B$), inflow and outflow ($IO$), retention time ($RT$) and evaporation ($Ev$). $Ev$ is the variable with the highest dispersion.

\begin{table}[htbp]
  \centering
  \caption{Physiochemical variables considered in the Arctic lake model.}
    {\footnotesize \begin{tabular}{|l|l|l|l|l|l|l|l|}
    \toprule
\toprule
    \textbf{Variable} & \textbf{Units} & \textbf{Acronym} & \textbf{Max} & \textbf{Min} & \textbf{Median} & \textbf{Mean}  & \textbf{std.\ dev.} \\
    \midrule
    Surface Light & MJ/m2/day & $SL$    & 30    & 1     & 5.1   & 11.06 & 11.27 \\
    Planktonic Ligth & MJ/m2/day & $PL$    & 28.2  & 1     & 4.9   & 10.46 & 10.57 \\
    Benthic Light & MJ/m2/day & $BL$    & 24.9  & 0.9   & 4.7   & 9.34  & 9.33 \\
    Surface Temperature & Deg C & $ST$    & 8.6   & 0     & 1.5   & 3.04  & 3.34 \\
    Planktonic Temperature & Deg C & $PT$    & 8.1   & 0.5   & 1.4   & 3.1   & 2.94 \\
    Benthic Temperature & Deg C & $BT$    & 7.6   & 1.6   & 2     & 3.5   & 2.29 \\
    Inflow and Outflow & m3/sec & $IO$  & 13.9  & 5.8   & 5.8   & 8.44  & 3.34 \\
    Retention Time & days  & $RT$    & 100   & 41.7  & 99.8  & 78.75 & 25.7 \\
    Evaporation & m3/day & $Ev$    & 14325 & 0     & 2436.4 & 5065.94 & 5573.99 \\
    Zone Mixing & \%/day & $ZM$    & 55    & 45    & 50    & 50    & 3.54 \\
    Inflow Conductivity & uS/cm & $ICd$   & 427   & 370.8 & 391.4 & 396.96 & 17.29 \\
    Planktonic Conductivity & uS/cm & $PCd$   & 650.1 & 547.6 & 567.1 & 585.25 & 38.55 \\
    Benthic Conductivity & uS/cm & $BCd$   & 668.4 & 560.7 & 580.4 & 600.32 & 40.84 \\
    Surface Oxygen & mg/litre & $SO2$   & 14.5  & 11.7  & 13.9  & 13.46 & 1.12 \\
    Planktonic Oxygen & mg/litre & $PO2$   & 13.1  & 10.5  & 12.6  & 12.15 & 1.02 \\
    Benthic Oxygen & mg/litre & $BO2$   & 13    & 9.4   & 12.5  & 11.62 & 1.51 \\
    Sediment Oxygen & mg/litre & $SdO2$ & 12.9  & 8.3   & 12.4  & 11.1  & 2.02 \\
    Inflow pH & ph Units & $IpH$   & 6.4   & 6     & 6.2   & 6.2   & 0.15 \\
    Planktonic pH & ph Units & $PpH$   & 6.7   & 6..40 & 6.6   & 6.57  & 0.09 \\
    Benthic pH & ph Units & $BpH$   & 6.6   & 6.4   & 6.5   & 6.52  & 0.07 \\
    \bottomrule
    \end{tabular}%
    }
  \label{tab:vars}%
\end{table}

Observing $RT$ and $IO$ in logarithmic scale, we can see that their values are located at the extremes, but their range is not long. Consequently, these variables have considerable variability in a short range. However, the ranges of the other variables do not reflect large changes.  This situation complicates the interpretation and comparison of the physiochemical dynamics. To attend this situation, we normalize the data to base $b$ of all points $x$ of all variables $X$ with the following equation:

\begin{equation}
f(x)=\left \lfloor b \cdot \frac{x-\min{X}}{\max{X}-\min{X}} \right \rfloor ,
\label{eq:norm}
\end{equation}
where $\lfloor x\rfloor$ is the floor function of $x$.

\begin{figure}[htbp]
\begin{center}
  \includegraphics[width=0.98\textwidth]{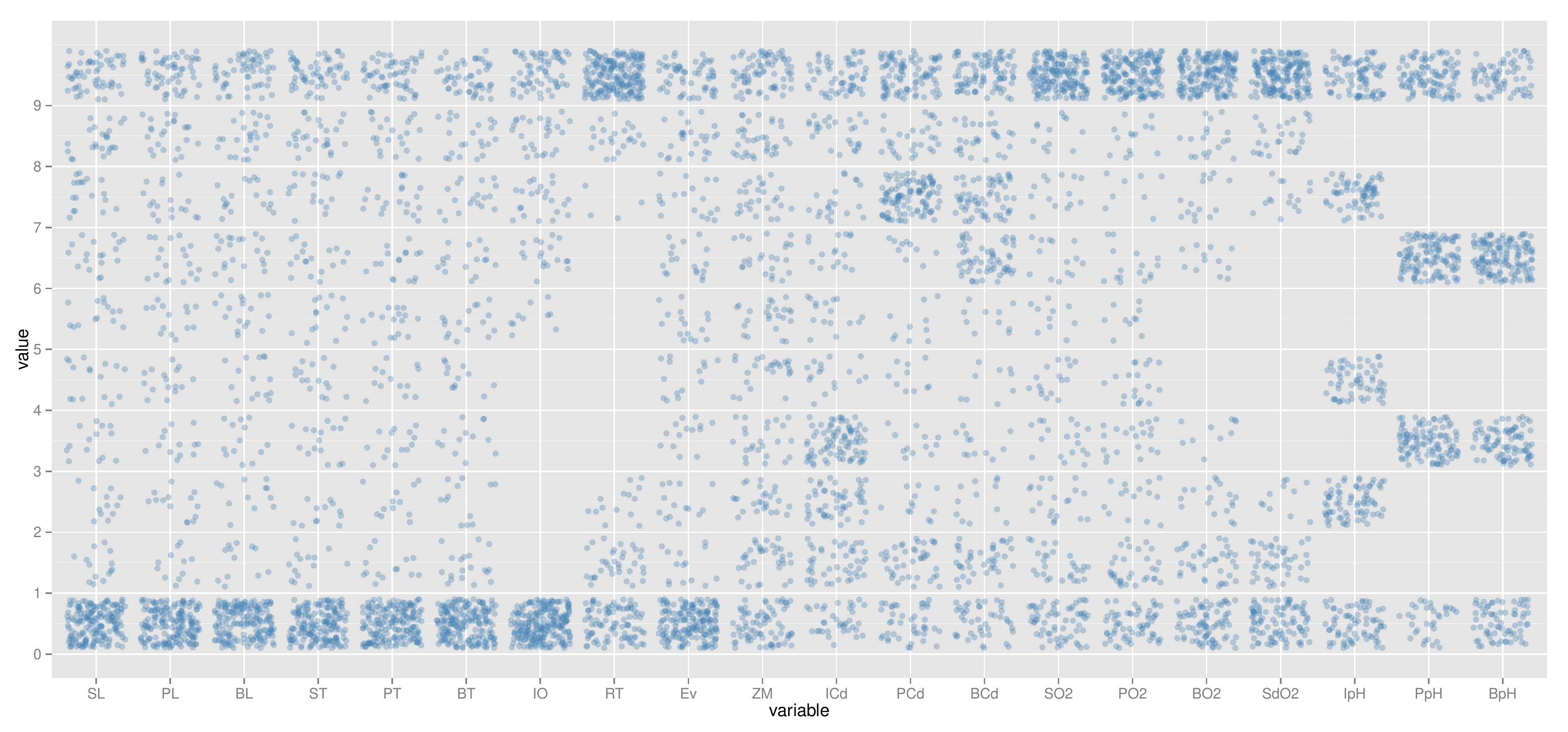}\\
\caption{Transformed variables from the physiochemical subsystem to base 10.}
\label{fig:norm}
\end{center}
\end{figure}

Once all variables are in transformed into a finite alphabet, in this case, base 10 ($b=10$), we can calculate emergence, self-organization, complexity, homeostasis and autopoiesis. Figure~\ref{fig:norm} depicts the number of points in each of the ten classes and shows the distribution of the values for each variable. Based on this distribution, the behavior for variables can be easily described and compared. Variables with a more homogeneous distribution will produce more information, yielding higher values of emergence. Variables with a more heterogeneous distribution will produce higher self-organization values.  The complexity of variables is not easy to deduce from Figure~\ref{fig:norm}.

\subsubsection{Emergence, Self-organization, and Complexity}

\begin{figure}[htbp]
\begin{center}
  \includegraphics[width=0.98\textwidth]{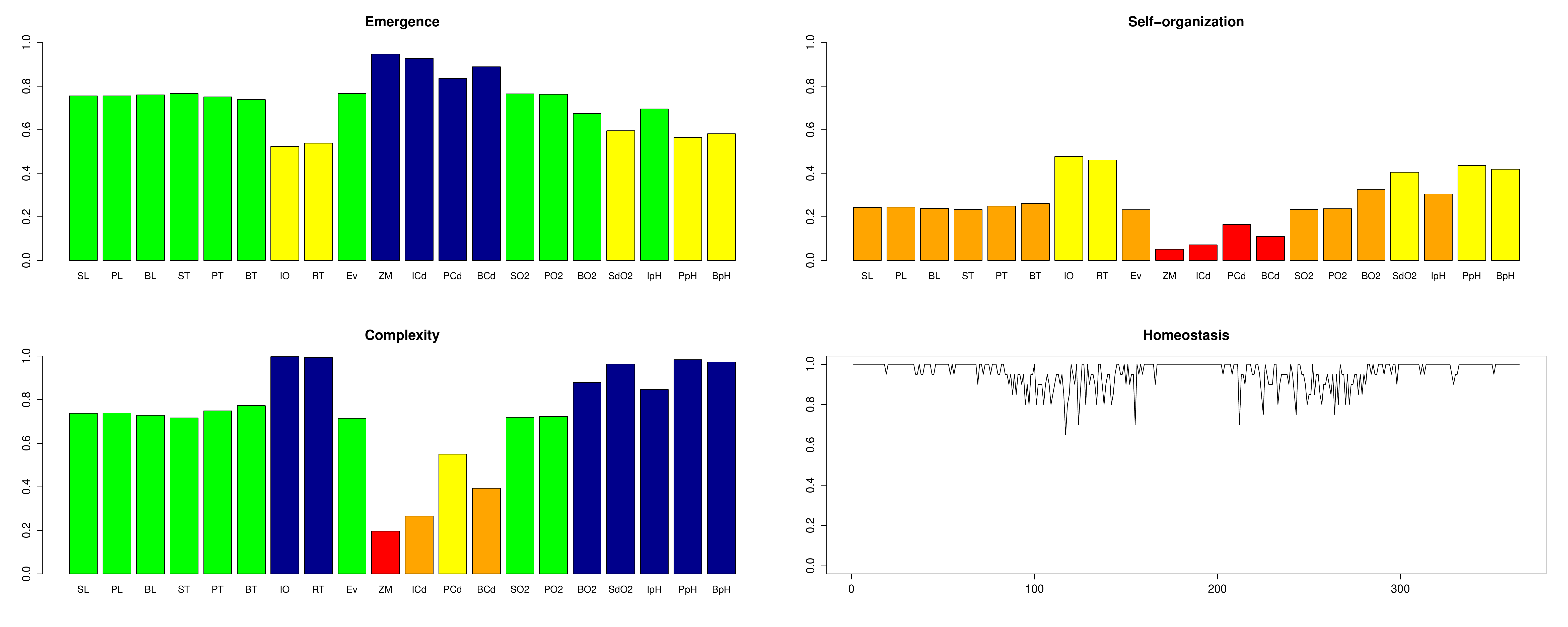}\\
\caption{$E$, $S$, and $C$ of physiochemical variables of the Arctic lake model (also shown in Table~\ref{tab:ESC}) and daily variations of homeostasis $H$ during a simulated year.}
\label{fig:ESCH}
\end{center}
\end{figure}

Figure~\ref{fig:ESCH} shows the values of emergence, self-organization, and complexity of the physiochemical subsystem. Variables with a high complexity $C \in [0.8,1]$ reflect a balance between change/chaos (emergence) and regularity/order (self-organization). This is the case of benthic and planktonic pH ($BpH$; $PpH$), $IO$ (Inflow and Outflow) and $RT$ (Retention Time). For variables with high emergencies ($E>0.92$), like Inflow Conductivity (ICd) and Zone Mixing ($ZM$), their change in time is constant; a necessary condition for exhibiting chaos. For the rest of the variables, self-organization values are low ($S<0.32$), reflecting low regularity. It is interesting to notice that in this system there are no variables with a high self-organization nor low emergence.

Since $E,S,C \in [0,1]$, these measures can be categorized into five categories as shown in Table \ref{tab:categories}. These categories are described on the basis of the range value, the color and the adjective in a scale from very high to very low. This categorization is inspired on the categories for Colombian water pollution indices. These indices were proposed by~\citet{Ramirez2003} and evaluated in~\citet{Fernandez2005}. 

\begin{table}[htbp]
  \centering
  \caption{Categories for classifying $E$, $S$, and $C$.}
    \begin{tabular}{|c|c|c|c|c|c|}
    \toprule
    \textbf{Category} & Very High & High  & Fair  & Low   & Very Low \\
    \midrule
    \textbf{Range} & $[0.8,1]$ & $[0.6,0.8)$ & $[0.4,0.6)$ & $[0.2,0.4)$ & $[0,0.2)$ \\
    \midrule
    \textbf{Color} & \cellcolor{blue!50} Blue  & \cellcolor{green!50} Green & \cellcolor{yellow!50} Yellow &  \cellcolor{orange!50} Orange & \cellcolor{red!50} Red \\
    \bottomrule
    \end{tabular}%
  \label{tab:categories}%
\end{table}

Table \ref{tab:ESC} shows results of $E$, $S$, and $C$ using the categories just mentioned.

\begin{table}[htbp]
  \centering
  \caption{$E$, $S$, and $C$ of physiochemical variables of the Arctic lake model. Also shown in Figure~\ref{fig:ESC}.}
    \begin{tabular}{|l|l|l|l|l|}
    \toprule
    \textbf{Variable} & \textbf{Acronym} & $E$ & $S$ & $C$ \\
    \midrule
    Benthic pH  & $BpH$   &\cellcolor{orange!50} 0.44196793 &\cellcolor{yellow!50}  0.55803207 & \cellcolor{blue!50} 0.98652912 \\
    In and Outflow  & $IO$  &\cellcolor{orange!50} 0.52310253 & \cellcolor{yellow!50} 0.47689747 & \cellcolor{blue!50} 0.99786509 \\
    Retention Time & $RT$    &\cellcolor{orange!50} 0.53890552 & \cellcolor{yellow!50} 0.46109448 &\cellcolor{blue!50} 0.99394544 \\
    Planktonic  pH & $PpH$   & \cellcolor{orange!50} 0.54122993 &\cellcolor{yellow!50}  0.45877007 &\cellcolor{blue!50} 0.99320037 \\
    Sediment Oxygen & $SdO2$  &\cellcolor{orange!50} 0.59328705 &\cellcolor{yellow!50}  0.40671295 & \cellcolor{blue!50} 0.96519011 \\
    Benthic Oxygen & $BO2$   & \cellcolor{green!50} 0.67904928 &\cellcolor{orange!50} 0.32095072 &\cellcolor{blue!50} 0.87176542 \\
    Inflow pH & $IpH$   & \cellcolor{green!50} 0.69570975 &\cellcolor{orange!50} 0.30429025 & \cellcolor{blue!50} 0.84679077 \\
    Benthic Temperature & $BT$    & \cellcolor{green!50} 0.72661539 &\cellcolor{orange!50} 0.27338461 &\cellcolor{green!50} 0.79458186 \\
    Planktonic Temperature & $PT$    &  \cellcolor{green!50} 0.75293885 &\cellcolor{orange!50} 0.24706115 &\cellcolor{green!50} 0.74408774 \\
    Planktonic  Light & $PL$    & \cellcolor{green!50} 0.75582978 &\cellcolor{orange!50} 0.24417022 &\cellcolor{green!50} 0.7382045 \\
    Surface Light & $SL$    & \cellcolor{green!50} 0.75591484 &\cellcolor{orange!50} 0.24408516 & \cellcolor{green!50} 0.73803038 \\
    Benthic Light & $BL$    & \cellcolor{green!50} 0.76306133 &\cellcolor{orange!50} 0.23693867 & \cellcolor{green!50} 0.72319494 \\
    Surface  Oxygen & $SO2$   &  \cellcolor{green!50} 0.76509182 &\cellcolor{orange!50} 0.23490818 & \cellcolor{green!50} 0.71890531 \\
    Surface  Temperature & $ST$    & \cellcolor{green!50} 0.76642121 &\cellcolor{orange!50} 0.23357879 & \cellcolor{green!50} 0.71607895 \\
    Evaporation & $Ev$    & \cellcolor{green!50} 0.76676234 &\cellcolor{orange!50} 0.23323766 & \cellcolor{green!50} 0.71535142 \\
    Planktonic Oxygen & $PO2$   & \cellcolor{green!50} 0.76887287 &\cellcolor{orange!50} 0.23112713 & \cellcolor{green!50} 0.71082953 \\
    Benthic  Conductivity  & $BCd$   & \cellcolor{green!50} 0.77974428 & \cellcolor{orange!50} 0.22025572 & \cellcolor{green!50} 0.68697255 \\
    Planktonic    Conductivity & $PCd$   & \cellcolor{green!50} 0.78604873 &\cellcolor{orange!50} 0.21395127 & \cellcolor{green!50} 0.6727045 \\
    Inflow  Conductivity & $ICd$   &\cellcolor{blue!50} 0.92845597 &\cellcolor{red!50} 0.07154403 & \cellcolor{orange!50} 0.26570192 \\
    Zone Mixing & $ZM$    &\cellcolor{blue!50} 0.94809050 &\cellcolor{red!50} 0.0519095 & \cellcolor{red!50} 0.1968596 \\
    \bottomrule
    \end{tabular}%
  \label{tab:ESC}%
\end{table}

From Table~\ref{tab:ESC} and a principal component analysis (not shown), we can divide the values obtained in complexity categories as follows:

\begin{description}
\item[Very High Complexity]: $C \in [0.8,1]$. The following variables balance self-organization and emergence: benthic and planktonic pH ($BpH$, $PpH$), inflow and outflow ($IO$), and retention time ($RT$). It is remarkable that the increasing of the hydrological regime during summer is related in an  inverse way with the dissolved oxygen ($SO2$; $BO2$). It means that an increased flow causes oxygen depletion.
Benthic Oxygen ($BO2$) and Inflow Ph ($IpH$) show the lowest levels of the category. Between both, there is a negative correlation: a doubling of $IpH$ is associated with a decline of $BO2$ in 40\%.  
  
\item[High Complexity]: $C \in [0.6,0.8)$. This group includes 11 of the 21 variables and involves a high $E$ and a low $S$. These 11 variables that showed more chaotic than ordered states are highly influenced by the solar radiation that defines the winter and summer seasons, as well as the hydrological cycle. These variables were: Oxygen ($PO2$, $SO2$); surface, planktonic and benthic temperature ($ST$, $PT$, $BT$); conductivity ($ICd$, $PCd$, $BCd$); planktonic and benthic light ($PL$,$BL$); and  evaporation ($Ev$).

\item[Very Low Complexity]: $C \in [0,0.2)$. In this group, $E$ is very high, and $S$ is very low.  This category includes the inflow conductivity ($ICd$) and water mixing variance ($ZM$). Both are high and directly correlated; it means that an increase of the mixing percentage between planktonic and benthic zones is associated with an increase of inflow conductivity.
\end{description}

\subsubsection{Homeostasis}

The homeostasis was calculated by comparing the daily values of all variables, representing the state of the Arctic subsystem. The temporal timescale is very important, because $H$ can vary considerably if we compare states every minute or every month.

The $h$ values have a mean ($H$) of 0.95739726 and a standard deviation of 0.064850247. The minimum $h$ is 0.60 and the maximum $h$ is 1.0. In an annual cycle, homeostasis shows four different patterns, as shown in Figure~\ref{fig:ESCH}, which correspond with the seasonal variations between winter and summer. These four periods show scattered values of homeostasis as the result of transitions between winter and summer and winter back again. The winter period (first and last days of the year) has very high $h$ levels (1 or close to 1) and starts from day 212 and goes to day 87. In this period, the winter conditions such as low light level, temperature, maximum time retention due to ice covering, low inflow and outflow, water mixing interchange between planktonic and benthic zones, low conductivities and pHs and high oxygen are present. The second, third and fourth periods correspond to summer. The second period starts with an increase of benthic pH, zone mixing, and inflow-outflow variables. Between days 83 and 154, this period is characterized for extreme fluctuations as a result of an increase in temperature and light. Homeostasis fluctuates and reaches a minimum of 0.6 in day 116. At the end of this period, the evaporation and zone mixing increase, while oxygen decreases in the benthic zone and sediment. The third period (days 155 to 162) reflects the stabilization of the summer conditions; It means maximum evaporation, temperature, light, mixing zone, conductivity and pH and the lowest oxygen level. Homeostasis is maximal again for this period. The fourth period (days 163-211), which has $h$ fluctuations near 0.9, corresponds to the transition of summer to winter conditions. 

As it can be seen, using $h$, periodic or seasonal dynamics can be followed and studied.

\subsubsection{Autopoiesis}

Autopoiesis was measured for three components (subsystems) at the planktonic and benthic zones of the Arctic lake. These were physiochemical, limiting nutrients and biomass. They include the variables and organisms related in Table~\ref{tab:comps}.

\begin{longtable}{|p{.17\linewidth}|p{.34\linewidth}|p{.34\linewidth}|}
  \caption{Variables and organisms used for calculating autopoiesis.}  \label{tab:comps}\\\hline
    \textbf{Component} & \textbf{Planktonic zone} & \textbf{Benthic zone} \\\hline
    Physiochemical & Light, Temperature, Conductivity, Oxygen, pH. & Light, Temperature, Conductivity, Oxygen, Sediment Oxygen, pH. \\\hline
    Limiting Nutrients & Silicates, Nitrates, Phosphates, Carbon Dioxide. & Silicates, Nitrates, Phosphates, Carbon Dioxide. \\\hline
    Biomass & Diatoms, Cyanobacteria, Green Algae, Chlorophyta. & Diatoms, Cyanobacteria, Green Algae. \\\hline
\end{longtable}%

\begin{figure}[htbp]
\begin{center}
  \includegraphics[width=0.6\textwidth]{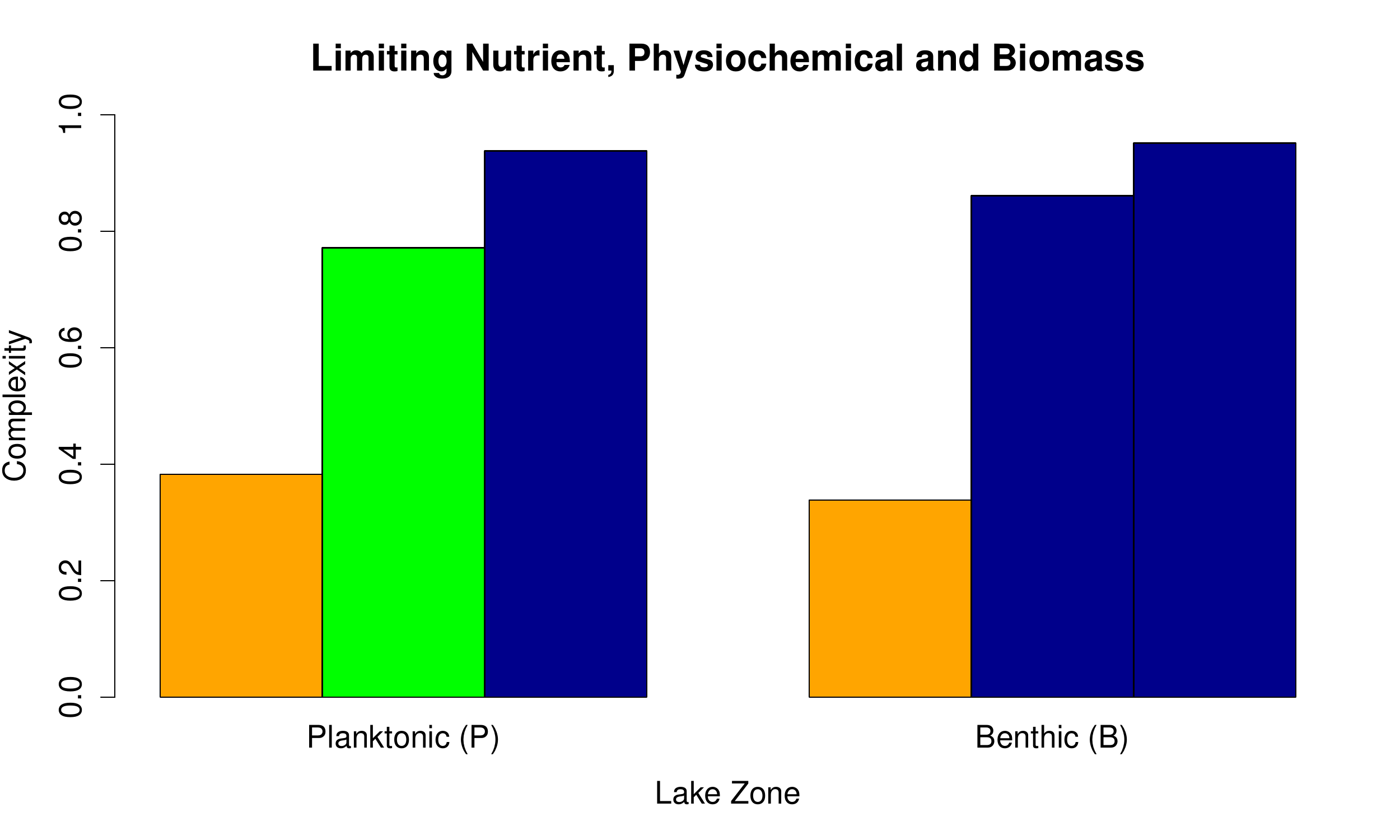}\\
\caption{$C$ of planktonic and benthic components.}
\label{fig:C}
\end{center}
\end{figure}

According to the complexity categories established in Table~\ref{tab:categories}, the planktonic and benthic components have been classified in the following categories: limiting nutrient variables in the low complexity category ($C\in[0.2,0.4)$; orange color), planktonic physiochemical variables in the high complexity category ($C\in[0.6,0.8)$; green color) and biomass and benthic physiochemical variables in the very high complexity category ($C\in[0.8,1]$; blue color). A comparison of the complexity level for each subsystem  of each zone (averaging their respective variables) is depicted in Figure~\ref{fig:C}.


\begin{figure}[htbp]
\begin{center}
  \includegraphics[width=0.6\textwidth]{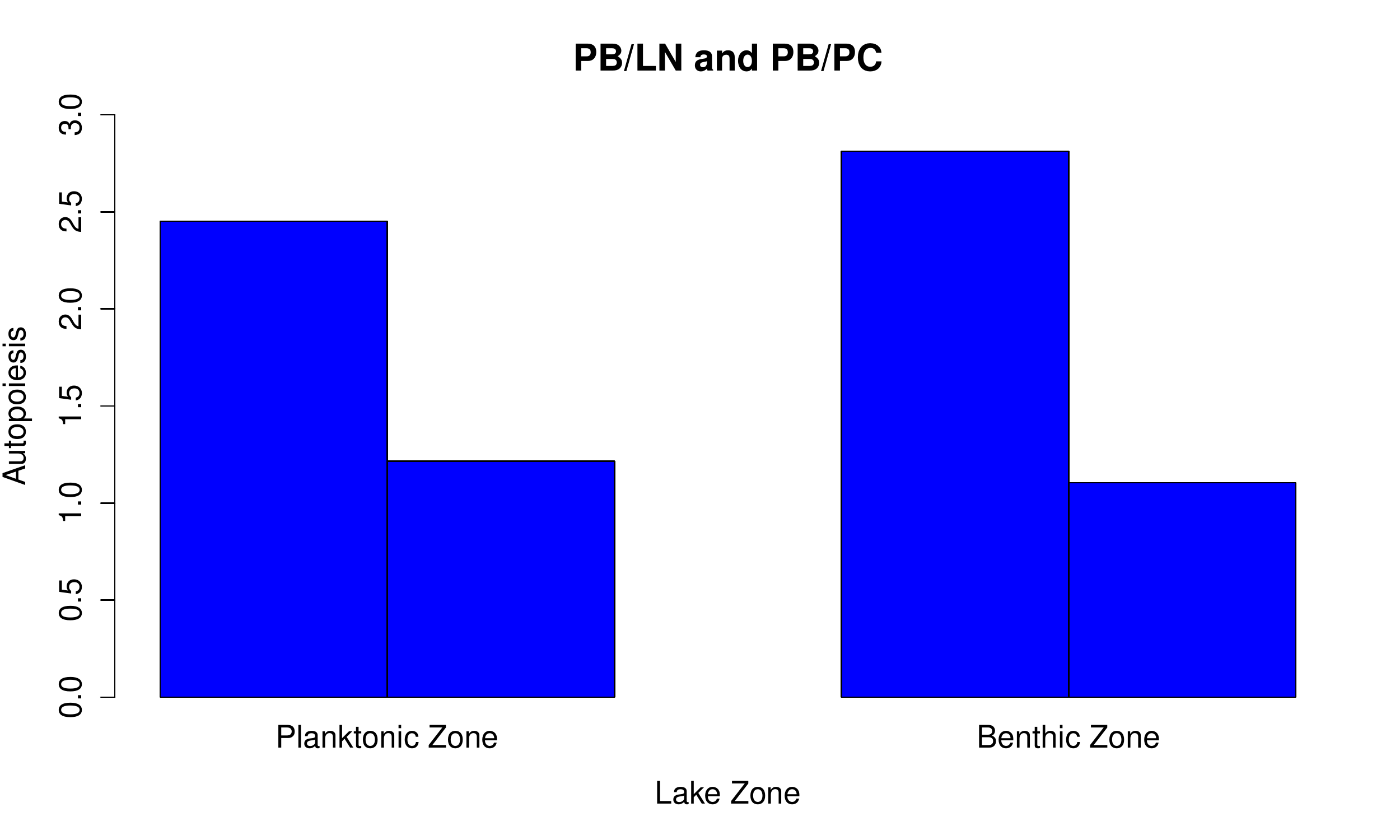}\\
\caption{$A$ of biomass depending on limiting nutrients and physiochemical components.}
\label{fig:A}
\end{center}
\end{figure}

In order to compare the autonomy of each group of variables, equation~\ref{eq:A} was applied to the complexity data, as shown in Figure~\ref{fig:A}. For the planktonic and benthic zones, we calculated the autopoiesis of the biomass elements in relation to limiting nutrient and physiochemical variables. All $A$ values are greater than 1. This means that the variables related to living systems have a greater complexity than the variables related to their environment, represented by the limiting nutrient and physiochemical variables. While we can say that some physiochemical variables, including limiting nutrients have more or fewer effects on the planktonic and benthic biomass, we can also estimate that planktonic and benthic biomass are more autonomous compared to their physiochemical and nutrient environments. The very high values of complexity of biomass imply that these living systems can adapt to the changes of their environments because of the balance between emergence and self-organization that they have.

\subsubsection{Multiple scales}

\begin{figure}
     \centering
     \subfigure[. Base 2.]{
          \label{fig:ESCH2}
          \includegraphics[width=.98\textwidth]{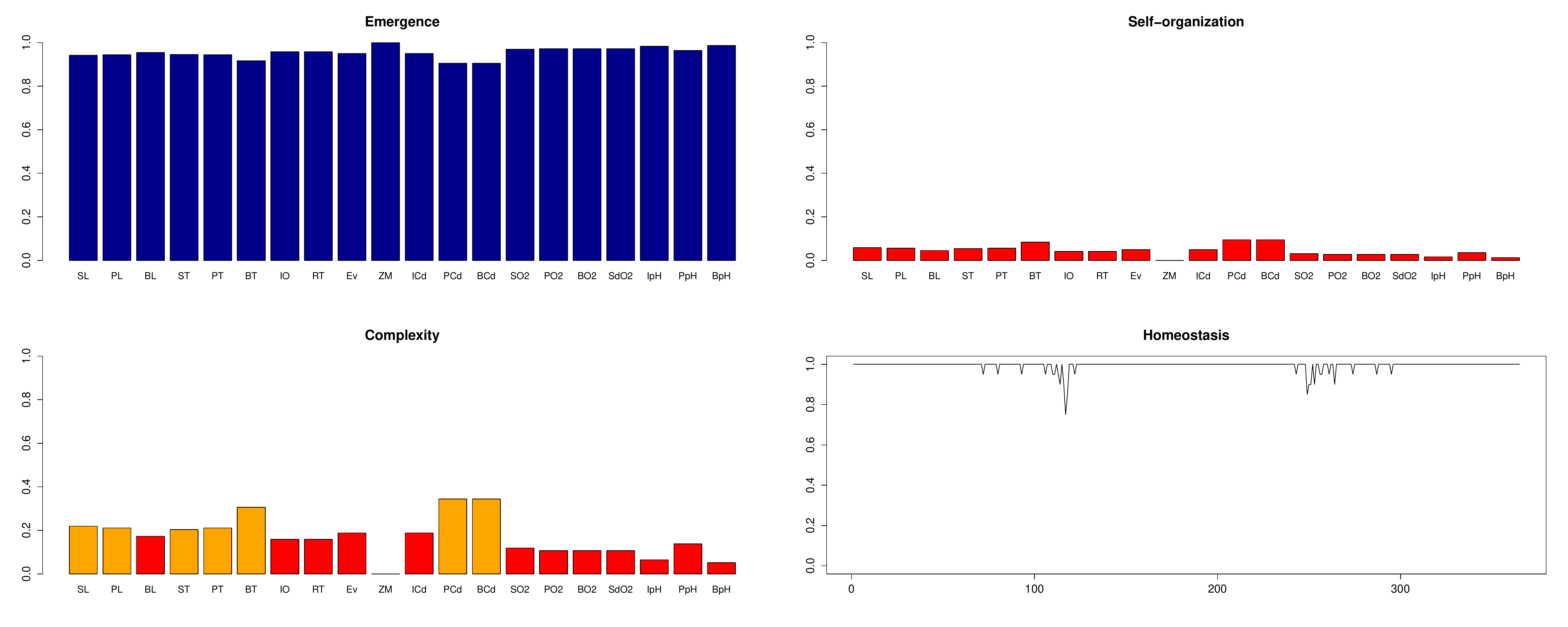}}
\\
     \subfigure[. Base 4.]{
          \label{fig:ESCH4}
          \includegraphics[width=.98\textwidth]{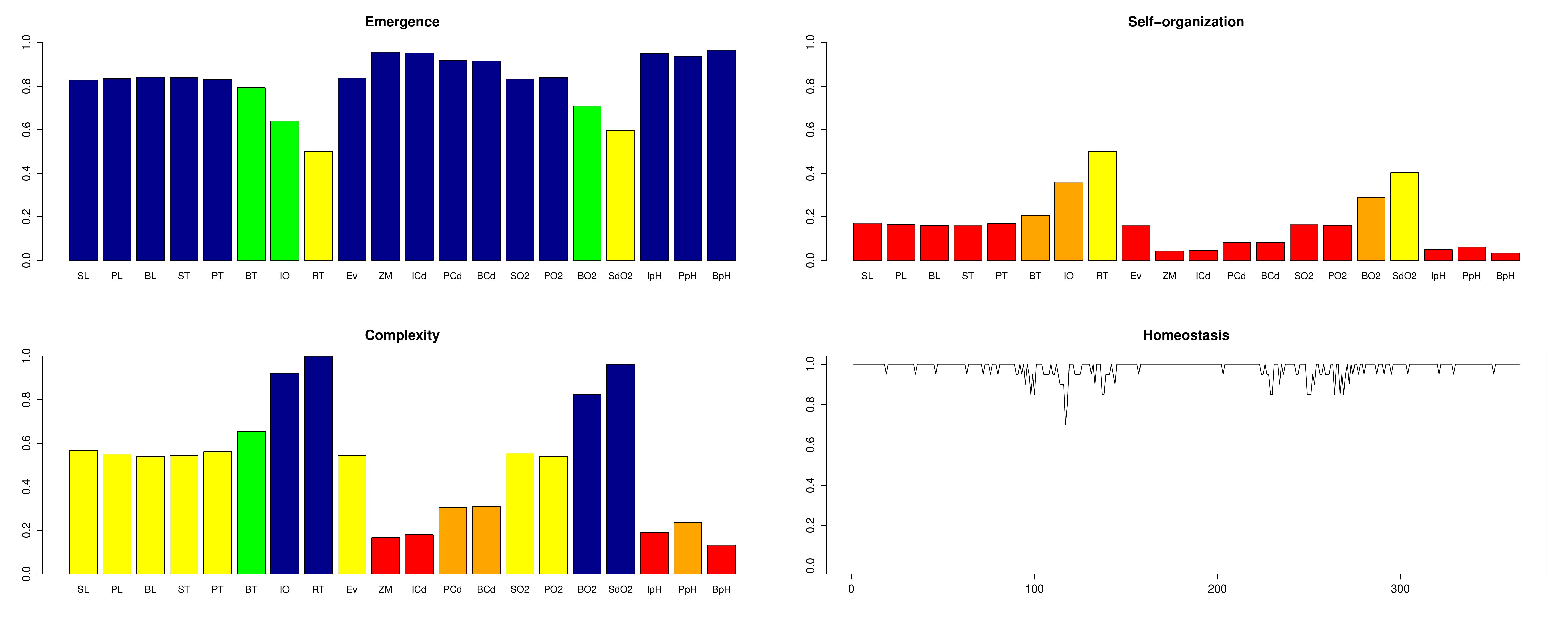}}
\\
     \subfigure[. Base 8.]{
          \label{fig:ESCH8}
          \includegraphics[width=.98\textwidth]{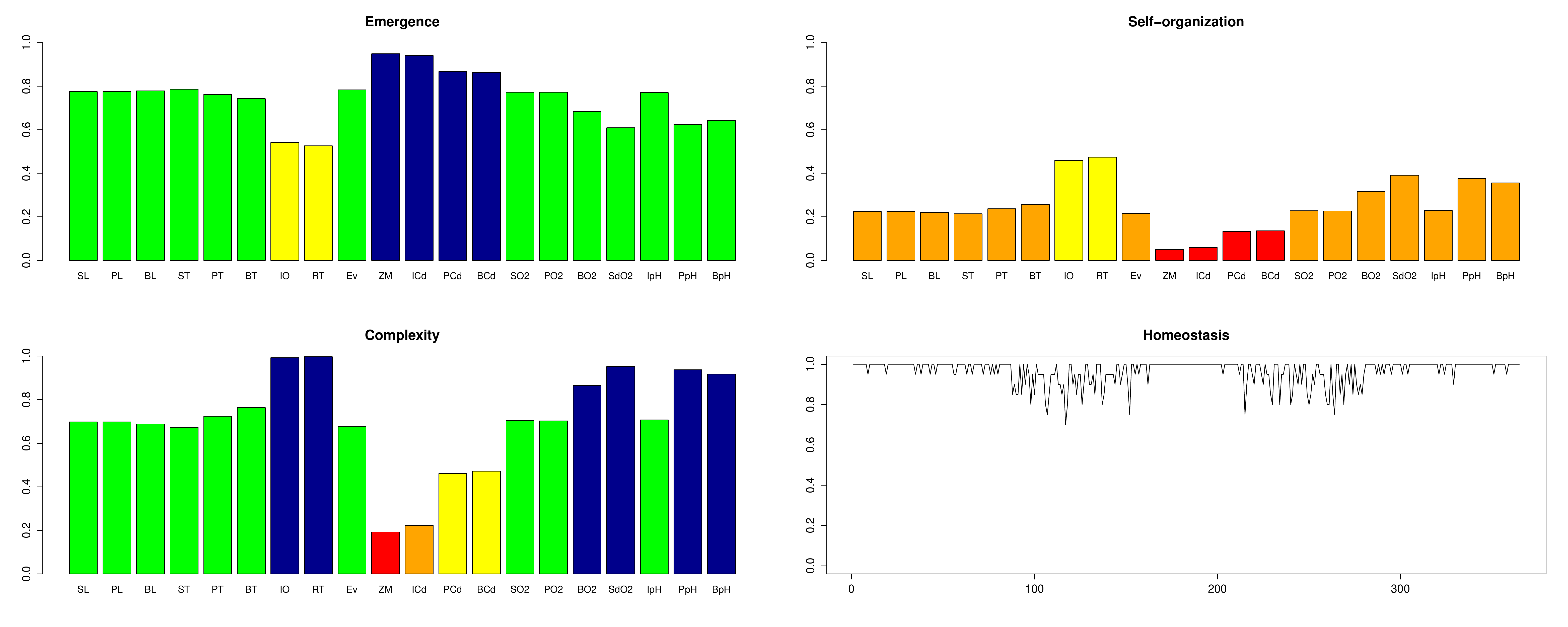}}
\\
     \caption{Emergence, Self-organization, Complexity, and Homeostasis for Arctic lake model at multiple scales: 2, 4, and 8.}
     \label{fig:ESCHbases1}
\end{figure}

\begin{figure}
     \centering
     \subfigure[. Base 16.]{
          \label{fig:ESCH16}
          \includegraphics[width=.98\textwidth]{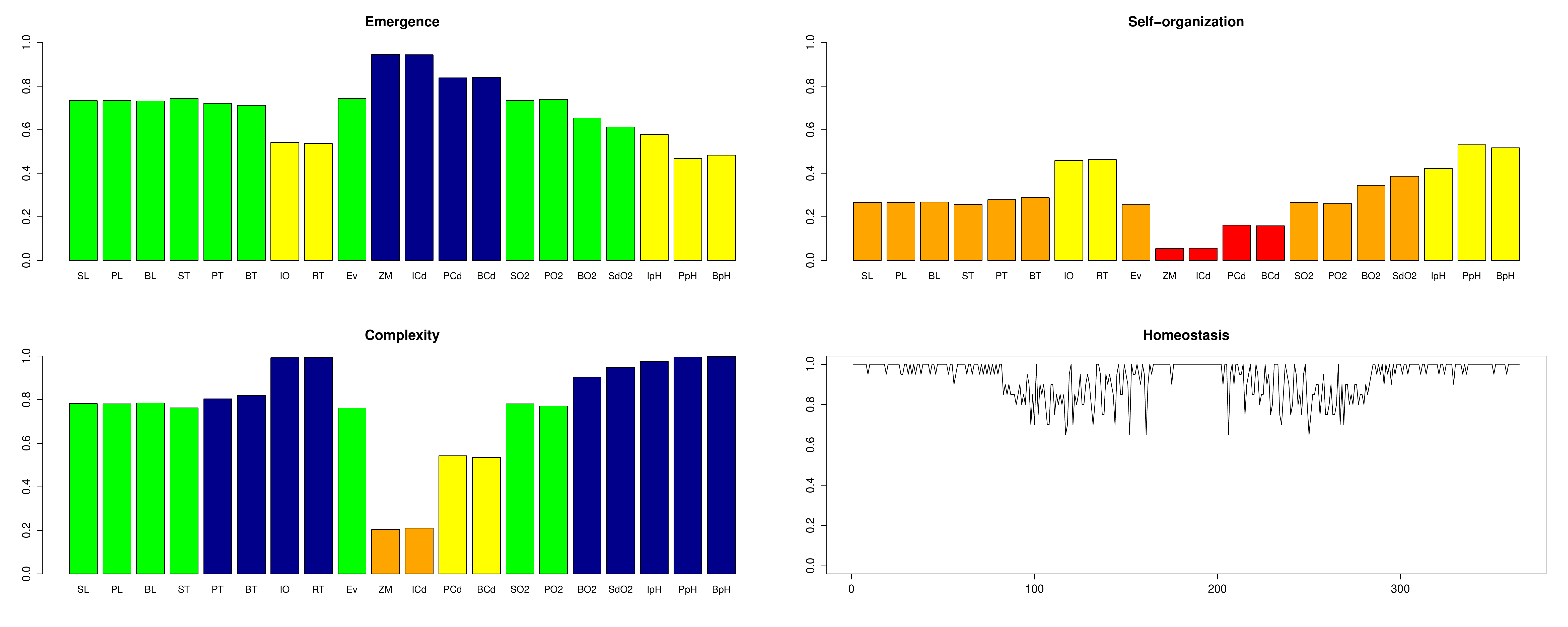}}
\\
     \subfigure[. Base 32.]{
          \label{fig:ESCH32}
          \includegraphics[width=.98\textwidth]{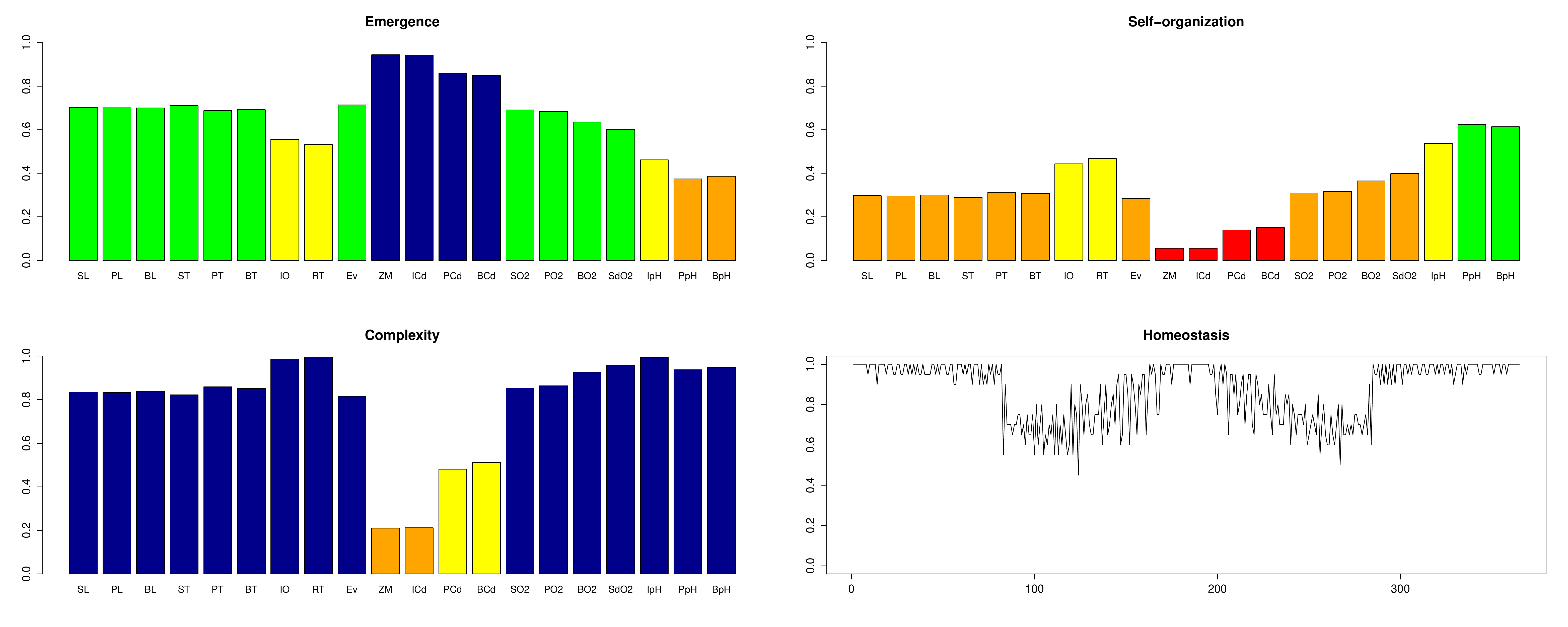}}
\\
     \subfigure[. Base 64.]{
          \label{fig:ESCH64}
          \includegraphics[width=.98\textwidth]{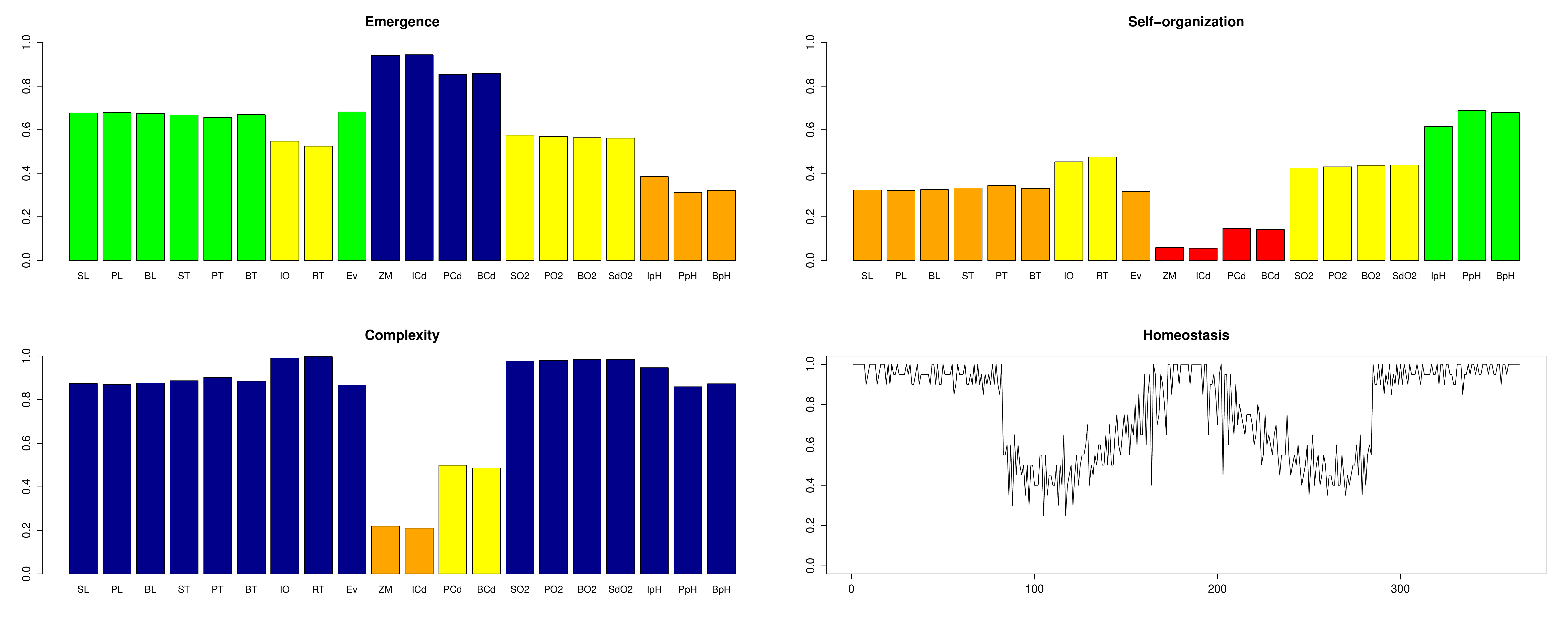}}
\\
     \caption{Emergence, Self-organization, Complexity, and Homeostasis for Arctic lake model at multiple scales: 16, 32, and 64.}
     \label{fig:ESCHbases2}
\end{figure}

The previous analysis of the Arctic lake was performed using base ten. We obtained the measures for the same data using bases $2^i,  \forall i \in [1..6]$, as shown in Figures~\ref{fig:ESCHbases1} and~\ref{fig:ESCHbases2}.

For base 2 (Figure~\ref{fig:ESCH2}), there is a very high $E$ for all variables, as the richness of the dynamics cannot be captured by only two values. Thus, $S$ and $C$ are low. Base 8 (Figure~\ref{fig:ESCH8}) gives results very similar to those of base 10 (Figure~\ref{fig:ESCH}), indicating that the measures are not sensitive to slight changes of base. Base 4 (Figure~\ref{fig:ESCH4}) gives intermediate values between base 2 and base 8. Results for bases 16, 32, and 64 (Figure~\ref{fig:ESCHbases2}) are very similar to those of base 10 and 8, showing that the choice of base is relevant but not a sensitive parameter. 

As more diversity is possible with higher bases, homeostasis values decrease with base. Still, the different periods of the year can be identified at all scales, with different levels of detail.

In the case of the Arctic lake model, studying the dynamics with a single base, i.e. at a single scale, can be very informative. However, studying the same phenomena at multiple scales can give further insights, independently on whether the measures change or not with scale.

\section{Discussion}

\subsection{Measures}

The proposed measures characterize the different configurations and dynamics that elements of complex systems acquire through their interactions.
Just like temperature averages the kinetic energy of molecules, much information is lost in the averaging, as the description of phenomena changes scale. The measures are probabilistic (except for $H$) and they all rely on statistical samples\footnote{This is also the reason for why all measures are unitless.}. Thus, the caveats of statistics and probability should be taken into consideration when using the proposed measures.
Still, these measures capture the properties and tendencies \emph{of a system}, that is why the scale at which they are described is important. They will not indicate which element interacted with which element, how and when. If we are interested in the properties and tendencies of the elements, we can change scale and apply the measures there. Still, we have to be aware that the measures are averaging---and thus simplifying---the phenomena they describe. Whether \emph{relevant} information is lost on the averaging depends not only on the phenomenon, but on what kind of information we are interested in, i.e.\ relevance is also partially dependent on the observer~\citep{Gershenson2002ua}.

\subsection{Complexity as balance or entropy?}

Some approaches relate complexity with a high entropy, i.e.\ information content~\citep{BarYam2004,Delahaye:2007}. Just as chaos should not be confused with complexity~\citep{Gershenson:2011e}, a very high entropy (high emergence $E$) implies too much change, where complex patterns are destroyed. On the other hand, very low entropy (high self-organization $S$), prevents complex patterns from emerging. As it has been proposed by several authors, complexity can be seen as balance between order and disorder~\citep{Langton1990,Kauffman1993,LopezRuiz:1995}, and thus, it is logical to postulate $C$ as a balance of $E$ and $S$.

It might seem contradictory to define emergence as the opposite of self-organization, as they are both present in several complex phenomena. However, when one takes one to the extreme (emergence or self-organization), the other is negligible. It is precisely when both of them are balanced that complexity occurs, but this does not mean that both of them have to be maximal.

\subsection{Fisher Information}

$C$ is correlated with Fisher information, which has been shown to be related to phase transitions~\citep{Prokopenko2011Relating-Fisher}. Following the view of high complexity as a balance, it is natural that $C$ is maximal at phase transitions, which is the case for both $C$ and Fisher information. However, the steepness of Fisher information is much higher than that of $C$. It is appropriate for determining phase transitions, but it makes little distinction of dynamics farther from transitions. $C$ is smoother, so it can represent dynamical change in a more gradual fashion. Moreover, to calculate Fisher information, a parameter must be varied, which limits its applicability for analyzing real data. This is because in many cases the data available is for a fixed set of parameters, with no variation. Under these circumstances, Fisher information cannot be calculated.

\subsection{Tsallis entropy}

Tsallis \citeyearpar{Tsallis1988} proposed a generalized measure of Shannon's information for non-ergodic systems. This measure has been correlated with complexity~\citep{Tsallis2002,GellManTsallis2004}. On the one hand, it would be interesting to compare Tsallis entropy with $C$ for different systems. On the other hand, it would be worth exploring what occurs when  $I$ is replaced with Tsallis entropy in $E$ (eq.~\ref{eq:E}) and how this affects $S$, $C$, and $A$ at multiple scales. 

\subsection{Guided Self-organization}

The measures proposed have several implications for GSO, beyond providing a measure of self-organization. In order to guide a complex system, one has to detect what kind of dynamical regime it has. Depending on this, and on the desired configuration for the system, different interventions can be made~\citep{Gershenson:2010}. The measures can inform directly about the dynamical regime and about the effect of the intervention. 

For example, if we want to have a system with a high complexity, first we need to measure what is its actual complexity. If it is not the desired one, then the dynamics can be guided. But we also have to measure the complexity during the guiding process, to evaluate the effectiveness of the intervention.


\subsection{Scales}

The proposed measures can be applied at different scales, with drastic outcomes. For example, the string '$1010101010$' will have $E=1$ in base 2, as $P(0)=P(1)=0.5$. However, in base 4, each symbol pair is transformed into a single symbol, so the string is transformed to '$22222$', and thus $P(2)=1$ and $P(0)=P(1)=P(3)=0$, giving $E=0$. Which scale(s) should be used is a question that has to be decided and justified. Multiscale profiles can be helpful in visualizing how the measures change with scale.


\subsection{Normalization}

For treating continuous data, we used equation~\ref{eq:norm} to normalize to a finite alphabet, which is equally distributed. Clustering methods could also be used to process data into finite categories. Still, an issue might arise for either case: if the available data does not represent the total range of possible values of a variable, e.g. data $\in[4.5,5.5]$ but the variable $\in[0,10]$. If we consider $b=10$, then equation~\ref{eq:norm} would produce ten categories for the available data, which might be homogeneously distributed and this give a high $E$. However, if we considered the variable range for equation~\ref{eq:norm}, it would categorize the available data in only two categories, leading to a low $E$. This problem is similar to the one of scales. We suggest to use the viability zone of a variable when known to normalize variables.

\subsection{Autopoiesis and Requisite Variety}
\label{sec:ReqVar}

Ashby's Law of Requisite Variety~\citep{Ashby1956} states that an active controller requires as much variety (number of states) as that of the controlled system to be stable. For example, if a system can be in four different states, its controller must be able to discriminate between those four states in order to regulate the dynamics of the system. 

The proposed measure of autopoiesis is related to the law of requisite variety, as a system with a $A>1$ must have a higher complexity (variety) than its environment, also reflecting its autonomy. Thus, a successful controller should have $A>1$ (at multiple scales~\citep{Gershenson:2010a}), although the controller will be more efficient if $A\rightarrow1$, assuming that higher complexities have higher costs.

\section{Conclusions}

We reviewed measures of emergence, self-organization, complexity, homeostasis, and autopoiesis based on information theory. Axioms were postulated for each measure and equations were derived form them. Having in mind that there are several different measures already proposed~\citep{ProkopenkoEtAl2007,GershensonFernandez:2012}, this approach allows us to evaluate the axioms underlying the measures, as opposed to trying to compare different measures without a common ground. 

The generality and usefulness of the proposed measures will be evaluated gradually, as these are applied to different systems. These can be abstract (e.g.\ Turing machines~\citep{Delahaye:2007,Delahaye:2012}, $\epsilon$-machines~\citep{ShaliziCrutchfield2001,GoernerupCrutchfield2008}), biological (ecosystems, organisms), economic, social or technological~\citep{Helbing:2011}.

The potential benefits of general measures as the ones proposed here are manifold. Even if with time more appropriate measures are found, aiming at the goal of finding general measures which can characterize complexity, emergence, self-organization, homeostasis, autopoiesis, and related concepts for any observable system is a necessary step to take.

\section*{Acknowledgements}

We should like to\ thank Nihat Ay, Ragnar Behncke, Paul Bourgine, Niccolo Capanni, Wilfried Elmenreich, Tom Froese, Virgil Griffith, Joseph Lizier, Luis Miramontes-Hercog, Roberto Murcio, Oliver Obst, Daniel Polani, Mikhail Prokopenko, Robert Ulanowicz, Rosalind Wang, and an anonymous referee for interesting critiques, discussions, and suggestions. Diego Lizcano and Cristian Villate helped with programming tasks. Figure~\ref{fig:lake} was made by Alb\'an Blanco.
C. G. was partially supported by SNI membership 47907 of CONACyT, Mexico.

\bibliographystyle{cgg}

\bibliography{carlos,sos,RBN,complex,information,COG,traffic,eco}

\end{document}